\documentclass[10pt, final, journal, letterpaper, twocolumn]{IEEEtran}

\usepackage[dvips]{graphicx}
\usepackage{times}
\usepackage{cite}
\usepackage{amsmath}
\usepackage{array}
\usepackage{amssymb}

\usepackage{stfloats}
\usepackage{rotating,threeparttable,booktabs}
\usepackage{bm}
\usepackage{dcolumn,booktabs}
\usepackage{multirow}
\usepackage{graphicx}
\usepackage{subfigure}
\usepackage{xcolor}
\begin{document}

\title{\huge {Linear Precoding Designs for Amplify-and-Forward Multiuser Two-Way Relay Systems}}

\author{\IEEEauthorblockN{Rui~Wang, Meixia Tao, \IEEEmembership{Senior Member,~IEEE}} and 	
\IEEEauthorblockN{Yongwei Huang \IEEEmembership{Member,~IEEE}}
\thanks{R. Wang and M. Tao are with the Department of Electronic Engineering at Shanghai Jiao Tong University, Shanghai, 200240,
P. R. China. Emails:\{liouxingrui, mxtao\}@sjtu.edu.cn. Y. Huang is with the Department of Mathematics, Hong Kong Baptist University, Kowloon
Tong, Hong Kong. Email: huang@hkbu.edu.hk.}
\thanks{This work is supported by the Joint Research Fund for Overseas Chinese, Hong Kong and Macao Young Scholars under grant 61028001,
the NSF of China under grant 60902019, and the NCET Program under grant NCET-11-0331.}
\thanks{Part of this work was presented at GLOBECOM 2011.}
}
\maketitle
\vspace{-1cm}
\begin{abstract}
Two-way relaying can improve spectral efficiency in two-user cooperative communications. It also has great potential in multiuser systems.
A major problem of designing a multiuser two-way relay system (MU-TWRS)
is transceiver or precoding design to suppress co-channel interference.
This paper aims to study linear precoding designs
for a cellular MU-TWRS where a multi-antenna base station (BS)
conducts bi-directional communications with multiple mobile stations (MSs) via a multi-antenna relay station (RS)
with amplify-and-forward relay strategy.
The design goal is to optimize uplink performance, including total mean-square error (Total-MSE) and sum rate,
while maintaining
individual signal-to-interference-plus-noise ratio (SINR) requirement for downlink signals.
We show that the BS precoding design with the RS precoder fixed
can be converted to a standard second order cone programming (SOCP) and the optimal solution is obtained efficiently.
The RS precoding design with the BS precoder fixed, on the other hand,
is non-convex and we present an iterative algorithm to find a local optimal solution.
Then, the joint BS-RS precoding is obtained by solving the BS precoding and the RS precoding alternately.
Comprehensive simulation is conducted to demonstrate the effectiveness of the proposed precoding designs.
\end{abstract}


\section{Introduction}
Due to complex wireless propagation environments, such as multi-path fading, shadowing and interference, the signals received by a remote destination receiver are not always strong enough to be decoded correctly. This problem has been considered as a main obstacle in the development of modern wireless communication systems. Recently, relay assisted cooperative communication has been proposed as an efficient way to deal with this problem, which now has received great attention from both academia and industry.
One example of the relay assisted cooperative communication is one-way relay system, which has been well studied in past decade \cite{Tang2007,Rong2009}. Although it has shown great potential in for example, transmission reliability, energy saving and coverage extension, one-way relaying on the other hand reduces spectral efficiency due to half-duplex constraint.

A promising technique to improve spectral efficiency of one-way relaying is to apply network coding \cite{Ahlswede2000},
resulting in
two-way relaying which has now attracted great attention \cite{Rankov2007,Zhang2009,RuiWang,Yuan2010}. Two-way relaying applies the principle of network coding at the relay node so as to mix the signals received from the two source nodes
who wish to exchange information with each other
and then employs at each destination self-interference (SI) cancelation to extract the desired information.
Compared with traditional one-way relaying, spectral efficiency of two-way relaying can be significantly improved since only two time slots  instead of four time slots are needed to complete one round of information exchange.

In this work, we consider two-way relaying in multiuser systems.
As in traditional multiuser systems, it is crucial to mitigate co-channel interference (CCI) for multiuser two-way relay system (MU-TWRS).
An advanced method to suppress CCI is to apply
multiple-input multiple-output (MIMO) technique.
Therein, transceiver or precoding should be carefully designed at each multi-antenna station, especially at the relay station (RS) \cite{Esli2008a,Esli2008,Joung2010,Joung2010a,Yilmaz2010, Leow2011, Tao2012, Ding2011}.
In \cite{Esli2008a,Esli2008}, authors study linear relay precoding
for MU-TWRS with decode-and-forward (DF) relay strategy.
Since the received signals is fully decoded in the first time slot,
the relay precoding only affects the transmission in the second time slot.
Then, by using
zero-forcing (ZF) precoding, the relay precoding studied in \cite{Esli2008a,Esli2008} reduces to
a power allocation problem.
The amplify-and-forward (AF) relay precoding, however, differs considerably from DF case as
the transmissions of the first and second time slots are tightly coupled and hence is more challenging.
Using ZF and minimum mean-square-error (MMSE) criteria,
authors in \cite{Joung2010,Joung2010a,Yilmaz2010,Leow2011} study precoding design for an AF based MU-TWRS with multiple pairs of users.
In particular, the explicit and analytical results are derived in \cite{Leow2011} for system performance evaluation.
Relay precoding design for the AF based MU-TWRS with multiple pairs of users is also considered in our previous work \cite{Tao2012}. Unlike \cite{Joung2010,Joung2010a,Yilmaz2010,Leow2011}, we do not impose any structural constraint on the relay precoder and thus the obtained results can approach the optimal performance \cite{Tao2012}.
In \cite{Ding2011}, authors study an AF MU-TWRS model with one base station (BS) and multiple mobile stations (MSs).
By using ZF precoding scheme,
explicit analytical results are also provided as in \cite{Leow2011}.
It is worth noting that the aforementioned ZF based precoding designs
all impose certain constraints on the number of relay antennas which may not be available for some scenarios.

In this paper, we consider linear precoding design for a cellular MU-TWRS where a multi-antenna BS intends to conduct bi-directional communications with multiple MSs via a multi-antenna RS.
Our work differs from \cite{Esli2008} in that we adopt AF relay strategy rather than DF for its simplicity in practical implementation. However, as mentioned previously, the precoding design with AF relay strategy is more challenging.
Our work is also different from \cite{Ding2011} since we do not impose any structures on precoders.
Our design goal is to enhance uplink performance
subject to individual signal-to-interference-plus-noise ratio (SINR) requirement for downlink signals.
Specifically, total mean-square error (Total-MSE) and sum rate are chosen to measure the performance of uplink.
Since linear precoding can be employed at the BS, RS or both, three associated optimization problems are considered.
When precoding is only conducted at the BS with the RS precoder fixed,
we show that this optimization problem can be converted to a standard second-order cone programming (SOCP), thus the optimal solution can be obtained efficiently.
The RS precoding with the BS precoder fixed, on the other hand, is non-convex
and we present an iterative algorithm to find a local optimal solution.
Thirdly, we obtain the joint BS-RS precoding design by solving the BS precoding and the RS precoding alternately, the convergence of which is guaranteed.
Simulation results show that the RS precoding scheme outperforms the BS precoding scheme in most cases and the joint precoding scheme outperforms the individual precoding scheme.
Besides performance, practical implementation issues, including signaling overhead and design complexity, for the proposed precoding designs are also discussed and compared.

The rest of the paper is organized as follows. In Section II, we present the system model.
Different precoding designs are presented in Section III.
In Section IV, we discuss the overhead and design complexity.
Extensive simulation results are illustrated in Section V. Finally, we conclude the paper in Section VI.

\emph{Notations}: $\cal E(\cdot)$ denotes the expectation over the
random variables within the brackets.
$\otimes$ denotes the Kronecker operator.
${\rm Tr}({\bf A})$, ${\bf A}^{-1}$, $\det({\bf A})$ and ${\rm Rank}(\bf A)$ stand for the trace, inverse, determinant and the rank of a matrix ${\bf A}$, respectively, and ${\rm Diag}(\bf a)$ denotes a diagonal matrix with ${\bf a}$ being its diagonal entries. Superscripts $(\cdot)^T$, $(\cdot)^{*}$ and $(\cdot)^H$ denote the transpose, conjugate and conjugate transpose, respectively.
${\bf 0}_{N\times M}$ implies the
$N\times M$ zero matrix.
${\bf I}_N$ denotes the $N \times N$ identity matrix and ${\bf I}_{N\times M}=[{\bf I}^T_M,{\bf 0}^T_{(N-M)\times M}]^T$ if $N\geq M$.
$||{\bf x}||^2_2$ denotes the squared Euclidean norm of a complex vector ${\bf x}$ and $||{\bf X}||^2_F$ denotes the Frobenius norm of a complex matrix ${\bf X}$. $|z|$ implies the norm of a complex number $z$, ${\Re}(z)$ and $\Im (z)$ denote its real and imaginary part, respectively. ${\mathbb C}^{x \times y}$ denotes the space of $x \times y$ matrices with complex entries. The distribution of a circular symmetric complex Gaussian vector with mean vector $\bf x$ and covariance matrix ${\bf \Sigma} $ is denoted by ${\cal CN}({\bf x},{\bf \Sigma})$.

\section{System Model}
Consider a multiuser two-way relay system where an $N$-antenna BS
conducts bi-directional communication
with $K$ single-antenna MSs
under the assistance of an $M$-antenna RS.
For effective multiuser transmission, we let $N \geq K$ and $M\geq K$.
Moreover, we assume that all the MSs are cell-edge users.
Thus, due to impairments such as multipath fading, shadowing and path loss of wireless channels,
the direct-path link between the BS and each MS is ignored. It is also assumed that the RS operates in half-duplex mode. That is, it cannot transmit and receive simultaneously.

\begin{figure}[t]
\begin{centering}
\includegraphics[scale=0.65]{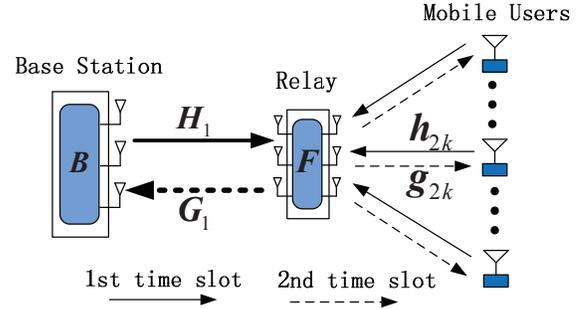}
\vspace{-0.1cm}
\caption{Illustration of a cellular MU-TWRS.} \label{fig:MacBc_two_way}
\end{centering}
\vspace{-0.3cm}
\end{figure}

The bi-directional (i.e., uplink and downlink) communications take place in two time slots as shown in Fig.~\ref{fig:MacBc_two_way}.
In the first time slot, also referred to as multiple-access (MAC) phase, both the BS and MSs simultaneously transmit their signals to the RS. The received $M\times1$ signal vector at the RS can be written as
\begin{equation}\label{eqnII-3}\nonumber
{\bf y}_R= {\bf H}_1 {\bf x}_B + \sum^K_{k=1} {\bf h}_{2k} s_k +{\bf n}_R,
\end{equation}
where ${\bf x}_B \in {\mathbb C}^{N \times 1}$ represents the transmit signal vector from the BS, ${s}_k$ denotes the transmit signal from the MS $k$.
We assume that the transmission power at the MS $k$ is $P_k$, i.e., ${\cal E}({s}_k{s}^*_k)=P_k$.
${\bf H}_1 \in {\mathbb C}^{M \times N}$ is the MIMO channel matrix from the BS to the RS, ${\bf h}_{2k}\in {\mathbb C}^{M \times 1}$ is the channel vector from the MS $k$ to the RS, and ${\bf n}_R$ denotes the additive noise vector at the RS following ${\cal CN}({\bf 0},\sigma^2_R{\bf I}_{M})$. Here ${\bf x}_B$ can be further expressed as
\begin{equation} \label{eqnII-1}\nonumber
{\bf x}_B={\bf B}{\bf s}_B,
\end{equation}
where ${\bf s}_B \in {\mathbb C}^{K\times 1}$ with ${\cal E}({\bf s}_B {\bf s}^H_B)={\bf I}_K$ is the modulated signal vector from the BS, ${\bf B}=\left[{\bf b}_1,{\bf b}_2,\cdots,{\bf b}_K\right] \in {\mathbb C}^{N\times K}$ denotes the transmit precoding matrix at the BS.
Furthermore, the maximum transmission power at the BS is assumed to be $P_B$, i.e.,
\begin{equation} \label{eqnII-2}
{\rm Tr}({\bf B}{\bf B}^H)\leq P_B.
\end{equation}

Upon receiving the superimposed signal ${\bf y}_R$, the RS performs linear processing by multiplying it with a precoding matrix $ {\bf F} \in {\mathbb C}^{M \times M}$
and then forwards it in the second time slot, also referred to as broadcast (BC) phase.
Therefore, the $M \times 1$ transmit signal vector from the RS is given by
\begin{equation} \label{eqnII-4}\nonumber
    {\bf x}_R={\bf F} {\bf y}_R= {\bf F}{\bf H}_1 {\bf x}_B + \sum^K_{k=1} {\bf F}{\bf h}_{2k} {s}_k + {\bf F}{\bf n}_R.
\end{equation}
The maximum transmission power at the RS is given by $P_R$, which yields
\begin{equation} \label{eqnII-5}
{\rm Tr}\left\{{\bf F}\left({\bf H}_1 {\bf B}{\bf B}^H {\bf H}^H_1 +{\bf H}_2{\bf P} {\bf P}^H{\bf H}^H_2 + \sigma^2_R {\bf I}_M
      \right){\bf F}^H\right\}\leq P_R,
\end{equation}
where we define ${\bf P}={\rm Diag}(\sqrt{P_1},\sqrt{P_2},\cdots,\sqrt{P_K})$ and ${\bf H}_2=\left[{\bf h}_{21},{\bf h}_{22},\dots,{\bf h}_{2K}\right]$.
Then the received signals at the BS and MS $k$ after the BC phase can be written as
\begin{equation} \label{eqnII-55}
\begin{split}
{\bf \tilde{y}}_B &= \sum^K_{k=1} {\bf G}_1{\bf F}{\bf h}_{2k} {s}_k + {\bf G}_1 {\bf F}{\bf H}_1{\bf B}{\bf s}_B +{\bf G}_1 {\bf F}{\bf n}_R + {\bf n}_B \\
     & = {\bf G}_1{\bf F}{\bf H}_2 {\bf s}_M + {\bf G}_1 {\bf F}{\bf H}_1{\bf B}{\bf s}_B + {\bf G}_1 {\bf F}{\bf n}_R + {\bf n}_B,
\end{split}
\end{equation}
\begin{equation} \label{eqnII-6}
\tilde{y}_k= \sum^K_{i=1}{\bf g}^T_{2k} {\bf F} {\bf H}_1 {\bf b}_i {s_{Bi}}
      +\sum^K_{i=1}{\bf g}^T_{2k} {\bf F}{\bf h}_{2i} {s}_i   + {\bf g}^T_{2k} {\bf F}{\bf n}_R + n_k.
\end{equation}
Here,  ${\bf s}_M=\left[s_1,s_2,\ldots,s_K\right]^T$,
$s_{Bi}$ denotes the $i$-th entry in ${\bf s}_B$,
${\bf G}_1 \in {\mathbb C}^{N\times M}$ and ${\bf g}_{2k}\in {\mathbb C}^{M\times 1}$ are the channel matrix and vector from the RS to the BS and MS $k$, respectively, ${\bf n}_B$ and $n_k$ denotes the additive noise at the BS and MS $k$, respectively, with ${\bf n}_B \sim {\cal CN}({\bf 0},\sigma^2_B{\bf I}_{N})$ and $n_k \sim {\cal CN}({0},\sigma^2_k)$. Note that both the BS and MS $k$ know their transmit signals ${\bf s}_B$ and $s_k$, respectively.
Therefore, the back propagated self-interference terms ${\bf s}_B$ and ${s}_k$ can be subtracted from \eqref{eqnII-55} and \eqref{eqnII-6}, respectively.
The equivalent received signals at the BS and MS $k$ are yielded, respectively, as
\begin{equation} \label{eqnII-7}
{\bf y}_B = {\bf G}_1{\bf F}{\bf H}_2 {\bf s}_M  + {\bf G}_1 {\bf F}{\bf n}_R + {\bf n}_B,
\end{equation}
\begin{equation} \label{eqnII-8}
\begin{split}
{y}_k &= {\bf g}^T_{2k} {\bf F} {\bf H}_1 {\bf b}_k {s_{Bk}} + \sum_{i\neq k}{\bf g}^T_{2k} {\bf F} {\bf H}_1 {\bf b}_i {s_{Bi}} \\
      & +\sum_{i\neq k}{\bf g}^T_{2k} {\bf F}{\bf h}_{2i} {s}_i
       + {\bf g}^T_{2k} {\bf F}{\bf n}_R + n_k.
\end{split}
\end{equation}

From \eqref{eqnII-8}, we find that the received downlink signal at each MS not only consists of the CCI from the downlink transmission (i.e., the second term), but also the CCI from the uplink transmission (i.e., the third term).
The downlink performance of each MS can be measured by SINR
given by
\begin{equation} \label{eqnII-9}
\begin{split}
      & \text{SINR}_k = \\
      &\frac{|{\bf g}^T_{2k} {\bf F}{\bf H}_1 {\bf b}_k|^2}
      {\sum_{l\neq k}\left(|{\bf g}^T_{2k} {\bf F}{\bf H}_1 {\bf b}_l|^2 +P_l |{\bf g}^T_{2k} {\bf F}{\bf h}_{2l}|^2 \right)
      +\sigma^2_R ||{\bf g}^T_{2k}{\bf F}||^2_2+\sigma^2_k }, \\
      & k=1,2,\cdots,K.
\end{split}
\end{equation}
As for the uplink transmission in \eqref{eqnII-7}, it can be viewed as a
MIMO multiple-access channel.
Depending on different performance requirements, various metrics can be used to evaluate its performance.
Our first objective aims to minimize the Total-MSE of all the MSs by assuming linear minimum mean-square error (MMSE) receiver at the BS.
Using Total-MSE for precoding design has been widely studied in multiuser systems \cite{Hunger2009,Joung2010,Joung2010a,Jorswieck2003,Luo2004}.
By minimizing MSE
\begin{equation} \label{eqnII-10-1}
e={\cal E}_{{\bf s}_M } \left(||{\bf W}{\bf y}_B-{\bf s}_M||^2_2 \right)
\end{equation}
with respect to the decoding matrix ${\bf W}$, the minimum Total-MSE is given by \cite{Palomar2003}
\begin{equation} \label{eqnII-10}
\begin{split}
e = {\rm Tr}\left( {\bf E}^{-1}\right),
\end{split}
\end{equation}
where
${\bf E} ={\bf I}_{K} + {\bf P}^H {\bf H}^H_2 {\bf F}^H  {\bf G}^H_1
(\sigma^2_R {\bf G}_1 {\bf F} {\bf F}^H {\bf G}^H_1 +\sigma^2_B {\bf I}_N )^{-1}
      {\bf G}_1 {\bf F} {\bf H}_2{\bf P}$
and the optimal ${\bf W}$ in \eqref{eqnII-10-1} is
\begin{equation} \label{eqnII-12}
\begin{split}
{\bf W} =& {\bf P}^H {\bf H}^H_2 {\bf F}^H  {\bf G}^H_1 \left( {\bf G}_1 {\bf F} {\bf H}_2 {\bf P}  {\bf P}^H {\bf H}^H_2 {\bf F}^H  {\bf G}^H_1  \right. \\
 & \left.  + \sigma^2_R {\bf G}_1 {\bf F} {\bf F}^H  {\bf G}^H_1 +\sigma^2_B {\bf I}_{N} \right)^{-1}.
\end{split}
\end{equation}
Our second objective aims
to maximize the sum rate of the uplink transmission. By applying successive interference cancelation (SIC) and linear MMSE filter at the BS,
the sum rate at the BS is given by \cite{Pramod2003}
\begin{equation} \label{eqnII-12-1}
\begin{split}
r = & 0.5 \log_2 \det \left({\bf I}_K + {\bf P}^H {\bf H}^H_2 {\bf F}^H  {\bf G}^H_1 \right. \\
& \left. (\sigma^2_R {\bf G}_1 {\bf F} {\bf F}^H {\bf G}^H_1 +\sigma^2_B {\bf I}_N )^{-1}
      {\bf G}_1 {\bf F} {\bf H}_2{\bf P}\right),
\end{split}
\end{equation}
where the factor $0.5$ is due to the fact that the MSs use two time slots to complete the uplink transmission. Note that
\eqref{eqnII-12-1} can be re-expressed as $r=0.5\log_2 \det({\bf E})$ with ${\bf E}$ defined in \eqref{eqnII-10}. We will see that the precoding designs proposed for Total-MSE minimization can be extended for sum rate maximization.

\section{Linear Precoding Designs}
From Section II, it is seen that
the downlink performance of each MS depends on both the BS precoder ${\bf B}$ and the RS precoder ${\bf F}$.
While for the uplink transmission,
it is only related to the RS precoder ${\bf F}$, thus less design freedom can be exploited
compared with the downlink.
In theory, the BS precoder $\bf B$ and the relay precoder $\bf F$ should be jointly designed such that the downlink and uplink performance can be optimized simultaneously. However, there is no single figure of merit to measure the overall performance of the multiuser bidirectional transmission. In this paper, we choose to ensure the downlink quality-of-service (QoS) for each individual MS while at the uplink minimizing the Total-MSE or maximizing the sum rate of all the users. This is because in practice the downlink data traffic usually is more dominant than the uplink traffic.
As such, the optimization problem is formulated as
%
\begin{eqnarray} \label{eqnII-13}
&&\min_{{\bf B}, {\bf F}} ~~e~ or~ -r \\ \nonumber
{s.t.} &&\text{SINR}_k \geq \lambda_k,~k=1,2,\cdots,K,~~\\ \nonumber
&& {\rm Tr}({\bf B}{\bf B}^H)\leq P_B \\ \nonumber
&& {\rm Tr}\left\{{\bf F}\left({\bf H}_1 {\bf B}{\bf B}^H {\bf H}^H_1 +{\bf H}_2{\bf P} {\bf P}^H{\bf H}^H_2 \right. \right. \\ \nonumber
   && \left. \left. + \sigma^2_R {\bf I}_M
      \right){\bf F}^H \right\}\leq P_R
\end{eqnarray}
where $\lambda_k$ is a preset threshold for the MS $k$.

Since linear precoding can be conducted at the BS, RS or both, three associated precoding designs are considered respectively in the following three subsections.
Note that for each design, the system needs different computational complexity and signaling overhead, such that they are suitable to different scenarios.

\subsection{BS precoding}
In this subsection, we assume that precoding is only employed at the BS, while the RS precoder is given as ${\bf F}=\alpha \tilde{{\bf F}}$ where $\tilde{{\bf F}}$ is an arbitrary fixed precoder applied at the RS, and $\alpha$ is a non-negative scalar used to scale the received signals at the RS to satisfy relay power constraint.
Note that besides maintaining the downlink SINR,
a properly designed ${\bf B}$
can reduce the RS power consumption by the signal ${\bf s}_B$ from the BS.
Then the uplink transmission can share more power at the RS,
which is helpful for improving its performance.


The optimization problem can be formulated as:
\begin{eqnarray} \label{eqnIII-1}
&&\min_{{\bf B}, \alpha} ~~f_1(\alpha)~~or~-f_2(\alpha)\\ \nonumber
 {s.t.}   && \rho_k  \geq \lambda_k, \forall k \\ \nonumber
&& {\rm Tr}({\bf B}{\bf B}^H)\leq P_B\\ \nonumber
&&  {\rm Tr}\left\{\alpha^2\tilde{{\bf F}} \left({\bf H}_1 {\bf B}{\bf B}^H {\bf H}^H_1 +{\bf H}_2{\bf P}{\bf P}^H{\bf H}^H_2 + \right. \right. \\ \nonumber
&& \left. \left. \sigma^2_R {\bf I}_{M}
      \right)\tilde{{\bf F}}^H \right\}\leq P_R \nonumber
\end{eqnarray}
where $f_1(\alpha) = {\rm Tr}\left({\bf E}(\alpha)^{-1} \right)$ and $f_2(\alpha) = \log_2 \det \left({\bf E}(\alpha) \right)$ with
${\bf E} (\alpha)={\bf I}_{K} + \alpha^2 {\bf P}^H {\bf H}^H_2 \tilde{{\bf F}}^H {\bf G}^H_1$
      $(\sigma^2_R \alpha^2 {\bf G}_1 \tilde{{\bf F}} \tilde{{\bf F}}^H {\bf G}^H_1 +\sigma^2_B {\bf I}_{N} )^{-1} {\bf G}_1 \tilde{{\bf F}}{\bf H}_2{\bf P}$
and
\begin{equation}\label{ADDSINR} \nonumber
\begin{split}
 \rho_k=
 \frac{{\alpha^2|{\bf g}^T_{2k} \tilde{{\bf F}}{\bf H}_1 {\bf b}_k|^2}}
     { \xi +\alpha^2 \sigma^2_R ||{\bf g}^T_{2k}\tilde{{\bf F}}||^2+\sigma^2_k },
\end{split}
\end{equation}
where $\xi = \sum_{l\neq k} \left(\alpha^2|{\bf g}^T_{2k} \tilde{{\bf F}}{\bf H}_1 {\bf b}_l|^2
     +\alpha^2 P_l|{\bf g}^T_{2k}\tilde{{\bf F}}{\bf h}_{2l}|^2 \right)$.
To proceed to solve \eqref{eqnIII-1}, we first give the following lemma, the proof of which is given in Appendix~\ref{prof_lemma1}.

\textbf{Lemma 1}:
$f_1(\alpha)$ and $-f_2(\alpha)$ are monotonically decreasing functions with respect of $\alpha$.

Based on \textit{Lemma 1}, it is easy to see that minimizing $f_1(\alpha)$ or $-f_2(\alpha)$ in \eqref{eqnIII-1} is equivalent to maximizing the scalar $\alpha$.
By defining ${\tilde{\bf B}}=\alpha{\bf B}$,
problem \eqref{eqnIII-1} can be re-expressed as:
\begin{eqnarray} \label{eqnIII-3}
      &&\max_{{\tilde{\bf B}}, \alpha} ~~ \alpha \\ \nonumber
      s.t. &&{\rm Tr}({\tilde{\bf  B}}{\tilde{\bf  B}}^H)\leq \alpha^2 P_B \\ \nonumber
      &&{\rm Tr}(\tilde{{\bf F}}{\bf H}_1 {\tilde{\bf  B}}{\tilde{\bf  B}}^H {\bf H}^H_1 \tilde{{\bf F}}^H ) \\ \nonumber
       && + \alpha^2 {\rm Tr}(  \tilde{{\bf F}}({\bf H}_2{\bf P}{\bf P}^H{\bf H}^H_2 + \sigma^2_R {\bf I}_{M}
      )\tilde{{\bf F}}^H)\leq P_R \\ \nonumber
      && \sum^K_{i=1}|{\bf g}^T_{2k} \tilde{{\bf F}} {\bf H}_1 {\tilde{\bf b}}_i|^2+
      \alpha^2 \left(\sum_{i\neq k} P_i |{\bf g}^T_{2k} \tilde{{\bf F}} {\bf h}_{2i}|^2+ \right. \\ \nonumber
      && \left. \sigma^2_R ||{\bf g}^T_{2k} \tilde{{\bf F}}||^2 \right) +\sigma^2_k
      \leq (1+\frac{1}{\lambda_k}) |{\bf g}^T_{2k} \tilde{{\bf F}} {\bf H}_1 {\tilde{\bf b}}_k|^2, \forall k
\end{eqnarray}

Although \eqref{eqnIII-3} is still a non-convex problem, we can use the observation made in \cite{Wiesel2006} that
any phase shift of ${\tilde{\bf b}}_k$, i.e., $e^{j\theta}{\tilde{\bf b}}_k$, does not affect the optimality of the primal problem. Therefore, for any optimal solutions, there always exists a phase shift version of ${\tilde{\bf b}}_k$ to make the term ${\bf g}^T_{2k}\tilde{{\bf F}} {\bf H}_1 {\tilde{\bf b}}_k$ real and positive while not affecting the value of the objective function and keeping the constraints satisfied. Thus, we can
convert problem \eqref{eqnIII-3} into the following equivalent form
\begin{eqnarray}\label{eqIII-4}
      && \max_{{\bf x}_B}~~ \alpha\\ \nonumber
      s.t. &&||{\tilde{\bf  B}}||^2_F\leq \alpha^2 P_B,\\ \nonumber
      && ||\tilde{{\bf F}} {\bf H}_1 {\tilde{\bf  B}}||^2_F \\ \nonumber
       && + \alpha^2 {\rm Tr}( \tilde{{\bf F}} ({\bf H}_2 {\bf P} {\bf P}^H{\bf H}^H_2 + \sigma^2_R {\bf I}_{M}
      )\tilde{{\bf F}}^H)\leq P_R\\ \nonumber
      && \sum^K_{i=1}|{\bf g}^T_{2k}\tilde{{\bf F}} {\bf H}_1 {\tilde{\bf b}}_i|^2+
      \alpha^2 \left(\sum_{i\neq k} P_i|{\bf g}^T_{2k} \tilde{{\bf F}} {\bf h}_{2i}|^2+  \right. \\ \nonumber
      && \left. \sigma^2_R ||{\bf g}^T_{2k}\tilde{{\bf F}}||^2_2 \right) +\sigma^2_k\leq (1+\frac{1}{\lambda_k}) (\underbrace{{\bf g}^T_{2k} \tilde{{\bf F}} {\bf H}_1 {\tilde{\bf b}}_k}
      _{real~and ~> 0}
      )^2, \forall k
\end{eqnarray}
where ${\bf x}_B=[vec({\tilde{\bf B}})^T, \alpha]$.
It is not hard to verify that \eqref{eqIII-4} is a standard second-order cone programming \cite{Convex} and the optimal solution can be obtained by using available software package \cite{CVX}. Then, dividing ${\tilde{\bf B}}$ by $\alpha$, we finally get the optimal ${\bf B}$.

\subsection{RS precoding}
In this subsection, we consider the precoding design at the RS with the BS precoder fixed.
In the following, we first consider the precoding design for Total-MSE minimization, then extend it to sum rate maximization.
\subsubsection{Total-MSE minimization}
The RS precoding to minimize Total-MSE can be formulated as:
\begin{eqnarray} \label{eqIII-B1}
      \min_{{\bf F}} &&
      {\rm Tr} \left( {\bf E}^{-1} \right) \\ \nonumber
      s.t. &&  \tau \leq P_R \\ \nonumber
      && \zeta_k \geq \lambda_k,~\forall k
\end{eqnarray}
where ${\bf E}$ is defined in \eqref{eqnII-10},
$\tau = {\rm Tr}\{{\bf F}({\bf H}_1 {\bf B}{\bf B}^H {\bf H}^H_1 +  {\bf H}_2{\bf P}{\bf P}^H{\bf H}^H_2 + \sigma^2_R {\bf I}_{M}
      ){\bf F}^H\}$ and
\begin{equation}\label{SINR} \nonumber
\begin{split}
& \zeta_k = \\
& \frac{|{\bf g}^T_{2k} {\bf F}{\bf H}_1 {\bf b}_k|^2}
      {\sum_{i\neq k}(|{\bf g}^T_{2k} {\bf F}{\bf H}_1 {\bf b}_i|^2 +P_i|{\bf g}^T_{2k} {\bf F}{\bf h}_{2i}|^2 )
      +\sigma^2_R ||{\bf g}^T_{2k}{\bf F}||^2_2+\sigma^2_k }.
      \end{split}
\end{equation}
Note that the power constraint at the BS is irrelevant here since ${\bf B}$ is fixed. It is not hard to verify that the objective function and $\text{SINR}$ constraints in \eqref{eqIII-B1} are both non-convex.
To make \eqref{eqIII-B1} more tractable,
we substitute the linear MMSE decoding matrix $\bf W$ back into \eqref{eqIII-B1} and rewrite it as:
\begin{eqnarray} \label{eqIII-B2}
      && \min_{{\bf F},{\bf W}} ~
      f({\bf F},{\bf W})\\  \nonumber
       s.t. && \tau \leq P_R \\  \nonumber
      && \zeta_k \geq \lambda_k,~\forall k
\end{eqnarray}
where
\begin{equation}\label{fFW}
\begin{split}
 f({\bf F},{\bf W})=& {\rm Tr}\left\{ {\bf W} {\bf G}_1 {\bf F} {\bf H}_2 {\bf P}  {\bf P}^H {\bf H}^H_2 {\bf F}^H  {\bf G}^H_1 {\bf W}^H + \right. \\
 & \left. \sigma^2_R {\bf W} {\bf G}_1 {\bf F} {\bf F}^H  {\bf G}^H_1 {\bf W}^H
      + \sigma^2_B {\bf W} {\bf W}^H + {\bf I}_{K}  \right.\\
     & \left. - {\bf W} {\bf G}_1 {\bf F} {\bf H}_2 {\bf P} -{\bf P}^H {\bf H}^H_2 {\bf F}^H  {\bf G}^H_1 {\bf W}^H \right\}.
\end{split}
\end{equation}
Note that \eqref{fFW} can also be computed from \eqref{eqnII-10-1}.
Although the two design matrices ${\bf W}$ and ${\bf F}$ are coupled together in \eqref{eqIII-B2}, the advantage of introducing $\bf W$ is that we can apply alternating optimization to solve two decoupled subproblems iteratively in what follows.

In the alternating optimization, the first step is to update the BS decoding matrix $\bf W$ for a given $\bf F$. From \eqref{eqIII-B2}, it is seen that the constraints are independent of ${\bf W}$. Thus,
the optimal $\bf W$ can be readily obtained as in \eqref{eqnII-12} by equating the gradient of the objective function in \eqref{eqIII-B2} to zero.

Secondly, we need to optimize $\bf F$ with $\bf W$ fixed.
This problem is equivalently rewritten as:
\begin{eqnarray} \label{eqIII-B4}
      \min_{{\bf F}} &&  {\rm Tr}\left\{  {\bf G}^H_1 {\bf W}^H {\bf W}  {\bf G}_1 {\bf F} \left( {\bf H}_2 {\bf P}  {\bf P}^H {\bf H}^H_2 \right. \right.\\ \nonumber
      && \left. + \sigma^2_R  {\bf I}_M \right) {\bf F}^H
      -  {\bf F} {\bf H}_2 {\bf P} {\bf W} {\bf G}_1  \\ \nonumber
      && \left. -{\bf G}^H_1 {\bf W}^H{\bf P}^H {\bf H}^H_2 {\bf F}^H   + \sigma^2_B {\bf W} {\bf W}^H + {\bf I}_{K}
       \right\} \\  \nonumber
      s.t.&& \tau \leq P_R \\  \nonumber
      && \zeta_k \geq \lambda_k,~\forall k
\end{eqnarray}
where we have used the fact that ${\rm Tr}({\bf A}{\bf B})={\rm Tr}({\bf B}{\bf A})$ for \eqref{fFW}.
Although we can verify that the objective function in \eqref{eqIII-B4} is convex based on \cite{Hjorungnes2007a}, while due to the non-convex $\text{SINR}$ constraints, the optimal $\bf F$ is still not easy to obtain.
To proceed, we need to recast \eqref{eqIII-B4} into a suitable form such that
efficient optimization tools can be applied. After
certain transformation as detailed
in Appendix~\ref{transformation}, problem \eqref{eqIII-B4} can be rewritten into the following inhomogeneous quadratically constrained quadratic program (QCQP) form \cite{Convex}:
\begin{subequations}\label{eqIII-B44}
\begin{align}
      \min_{{\bf f}} ~~& {\bf f}^H {\bf Q}_0 {\bf f} - {\bf f}^H {\bf q}_0 -{\bf q}^H_0 {\bf f} +q_0 \label{eqIII-B44-1}\\
      s.t. ~~& {\bf f}^H {\bf Q}_x {\bf f} \leq P_R \label{eqIII-B44-2}\\
     & {\bf f}^H {\bf Q}_k {\bf f} \geq \lambda_k \sigma^2_k,~ \forall k \label{eqIII-B44-3}
\end{align}
\end{subequations}
where ${\bf f}$, ${\bf Q}_0$, ${\bf Q}_x$ and ${\bf Q}_k$ are defined in \eqref{eqIII-B5}, \eqref{eqIII-B5-3} and \eqref{eqIII-B7} in Appendix~\ref{transformation}, respectively.
By checking the positive semidefiniteness of ${\bf Q}_0$ and the positive definiteness of ${\bf Q}_x$,
we can verify that both the objective function \eqref{eqIII-B44-1} and the RS power constraint \eqref{eqIII-B44-2} are convex. However,
the constraint \eqref{eqIII-B44-3} is not  concave due to that
${\bf Q}_k$ defined in  \eqref{eqIII-B7} is not necessarily negative semidefinite. Hence, optimization problem \eqref{eqIII-B44} is non-convex.
To solve \eqref{eqIII-B44}, we rewrite \eqref{eqIII-B44} into a standard QCQP form as follows:
\begin{eqnarray} \label{eqIII-B8}
      \min_{{\bf x}_F} && {\bf x}^H_F {\tilde{\bf Q}_0} {\bf x}_F\\ \nonumber
                      s.t. && |t|^2=1\\ \nonumber
     && {\bf x}^H_F {\tilde{\bf Q}_x} {\bf x}_F \leq 0 \\ \nonumber
   && {\bf x}^H_F {\tilde{\bf Q}_k} {\bf x}_F \leq 0, \forall k~
\end{eqnarray}
where ${\bf x}_F=[t,{\bf f}^T]^T$,
${\tilde{\bf Q}_0}=\begin{bmatrix}
                      q_0 & -{\bf q}^H_0 \\
                      -{\bf q}_0 & {\bf Q}_0
                     \end{bmatrix}$,
${\tilde{\bf Q}_x}=\begin{bmatrix}
                      -P_R & {\bf 0}_{1\times M^2} \\
                      {\bf 0}_{M^2\times 1} & {\bf Q}_x
                      \end{bmatrix}$ and
${\tilde{\bf Q}_k}=\begin{bmatrix}
                      \lambda_k \sigma^2_k & {\bf 0}_{1\times M^2} \\
                      {\bf 0}_{M^2\times 1} & -{\bf Q}_k
                      \end{bmatrix}$.
Note that \eqref{eqIII-B4} and \eqref{eqIII-B8} are equivalent to each other. If we get an optimal solution of \eqref{eqIII-B8}, we can always obtain  an optimal solution of \eqref{eqIII-B44} by selecting appropriate entries from ${{\bf x}_F}/{t}$ no matter $t$ is real or complex. By a close inspection of \eqref{eqIII-B8}, we find that \eqref{eqIII-B8} can be transformed into the following semidefinite programming (SDP) form \cite{Convex}:
\begin{eqnarray} \label{eqIII-B9}
      \min_{{\bf X}_F\succeq 0} &&  {\rm Tr}({\tilde{\bf Q}_0}{\bf X}_F)\\ \nonumber
      s.t.&& {\rm Rank}({\bf X}_F)=1,~
       \\ \nonumber
      &&  {\rm Tr}({\bf Q}{\bf X}_F)=1 \\ \nonumber
      && {\rm Tr}({\tilde{\bf Q}_x}{\bf X}_F)\leq 0 \\ \nonumber
      && {\rm Tr}({\tilde{\bf Q}_k}{\bf X}_F)\leq 0,  \forall k
\end{eqnarray}
where
${\bf Q}=\begin{bmatrix}
                      1 & {\bf 0}_{1\times M^2} \\
                      {\bf 0}_{M^2\times 1} & {\bf 0}_{M^2\times M^2}
                      \end{bmatrix}$.
Due to the rank-one constraint, it is not easy to obtian an optimal solution of \eqref{eqIII-B9}.
We therefore resort to relaxing it
by deleting the rank-one constraint, namely,
\begin{eqnarray} \label{eqIII-B10}
      \min_{{\bf X}_F\succeq 0} &&  {\rm Tr}({\tilde{\bf Q}_0}{\bf X}_F)\\ \nonumber
      s.t.
      && {\rm Tr}({\bf Q}{\bf X}_F)=1 \\ \nonumber
      && {\rm Tr}({\tilde{\bf Q}_x}{\bf X}_F)\leq 0 \\ \nonumber
      && {\rm Tr}({\tilde{\bf Q}_k}{\bf X}_F)\leq 0,  \forall k
\end{eqnarray}
Note that \eqref{eqIII-B10} is a standard SDP problem, thus its optimal solution can be easily obtained by using the available software package \cite{CVX}.
If the optimal solution of \eqref{eqIII-B10} is rank-one, the optimal RS precoder can be obtained by using eigenvalue decomposition. Otherwise, certain techniques are required to find the optimal RS precoder.

In what follows,
we first consider a system with no more than two MSs (i.e., $K\leq 2$)
for which an optimal solution of \eqref{eqIII-B44} can be obtained in most cases.
Then, we extend the results to a more general system with $K>2$ where the randomization technique is applied to find a quasi-optimal solution.

\paragraph{$K\leq 2$}
We first give the following theorem.

{\textbf {Theorem} 1}:
Suppose that the considered cellular MU-TWRS has at most two MSs, i.e., $K\leq 2$, an optimal rank-one solution of the non-convex optimization problem \eqref{eqIII-B9} can be derived in polynomial time from the relaxed SDP problem \eqref{eqIII-B10} in the following cases: 1) problem \eqref{eqIII-B10} has an optimal rank-one solution; 2) problem \eqref{eqIII-B10} has at least one inactive constraint at the optimal solution; 3) problem \eqref{eqIII-B10} has an optimal solution of rank higher than two if all the constraints are active.
\begin{proof}
Please refer to Appendix~\ref{prof_Theorem1}.
\end{proof}
From \textit{Theorem 1}, we find that we cannot obtain an optimal rank-one solution if the SDP relaxation problem \eqref{eqIII-B10} happens to have an optimal solution of rank two with all the constraints being active. However, our simulations show that this case has rarely occurred. Nonetheless, we can propose a procedure of producing a suboptimal rank-one solution in Appendix~\ref{procedure} for that special case.

Now, the iterative RS precoding algorithm to minimize Total-MSE for $K \leq 2$ can be outlined as follows.

\vspace{-0.2cm}
\hrulefill
\par
{\footnotesize
\textbf{Algorithm 1} (RS precoding with $K\leq 2$)
\begin{itemize}
\item \textbf{Initialize} ${\bf F}$ 
\item \textbf{Repeat}
\begin{itemize}
\item Update the BS decoding matrix ${\bf W}$ using \eqref{eqnII-12} for a fixed ${\bf F}$;
\item Update the RS precoder ${\bf F}$ with ${\bf W}$ fixed as follows: If the obtained ${\bf X}_F$ in \eqref{eqIII-B10} is rank-one, using eigenvalue decomposition to get ${\bf F}$. Otherwise, using the procedures presented in Appendix~\ref{prof_Theorem1} or~\ref{procedure} to get ${\bf F}$;
\end{itemize}
\item \textbf{Until} termination criterion is satisfied.
\end{itemize}}
\vspace{-0.3cm}
\hrulefill

\textbf{Lemma 2}: Algorithm 1 is convergent and the limit point of iteration is a stationary point of
\eqref{eqIII-B2}.
\begin{proof}
Since for $K\leq 2$, the optimal solution in \eqref{eqIII-B4} can be obtained in most cases as claimed in \textit{Theorem 1},
the solution in each iteration in Algorithm 1 can be viewed as being optimal.
Thus the Total-MSE at the BS is strictly reduced after each iteration before convergence.
On the other hand, the objective function is lower-bounded (at least zero).
Therefore, we conclude that Algorithm 1 is convergent.
We assume that the limit point of Algorithm 1 is $\left\{ \bar{\bf W}, \bar{\bf F}\right\}$.
At the limit point, the solution will not change if we continue the iteration. Otherwise, the Total-MSE can be further decreased and it  contradicts the assumption of convergence.
The optimal solution in each iteration
further means that $\bar{\bf W}$ and $\bar{\bf F}$ are local minimizers of each subproblem. Hence, we have
\begin{equation}\label{eqIII-B11}\nonumber
\begin{split}
    & {\rm Tr}\left\{{\triangledown}_{{\bf W}} {f}\left(\bar{\bf W}; \bar{\bf F}\right)^T \left( {\bf W}-\bar{\bf W}\right) \right\}\geq 0,\\
    & {\rm Tr}\left\{{\triangledown}_{{\bf F}} {f}\left(\bar{\bf F}; \bar{\bf W}\right)^T \left( {\bf F}-\bar{\bf F}\right) \right\}\geq 0,
\end{split}
\end{equation}
Summing up the two inequalities, we get
\begin{equation}\label{eqIII-B12}
    {\rm Tr}\left\{{\triangledown}_{{\bf X}} {f}\left(\bar{\bf X}\right)^T \left( {\bf X}-\bar{\bf X}\right) \right\}\geq 0,
\end{equation}
where ${\bf X}=\left[{\bf W}, {\bf F}\right]$. Condition \eqref{eqIII-B12} implies the stationarity of $\bar{\bf X}$ in \eqref{eqIII-B2} (e.g., see
\textit{Theorem 3} of \cite{Razaviyayn2012}).
\end{proof}

\paragraph{$K > 2$}
Now we consider a more general case with $K>2$.
Since at least five constraints are contained in \eqref{eqIII-B10},
it is difficult to find an optimal rank-one solution
if the optimal solution in \eqref{eqIII-B10} has higher rank than one. Next we propose to apply the randomization technique in \cite{Boyd2003} to find a quasi-optimal rank-one solution of \eqref{eqIII-B44}. We first transform \eqref{eqIII-B44} into the following equivalent form:
\begin{eqnarray}\label{eqIII-B13}
\min_{{\bf f}} && {\rm Tr}\left( {\bf Q}_0 {\tilde{\bf F}}\right) - {\bf f}^H {\bf q}_0 -{\bf q}^H_0 {\bf f} +q_0 \\ \nonumber
      s.t. && {\rm Tr}\left( {\bf Q}_x {\tilde{\bf F}}\right) \leq P_R \\ \nonumber
     && {\rm Tr}\left( {\bf Q}_k {\tilde{\bf F}}\right) \geq \lambda_k \sigma^2_k, \forall k \\ \nonumber
     && {\tilde{\bf F}}= {\bf f} \times {\bf f}^H
\end{eqnarray}
Relaxing the constraint ${\tilde{\bf F}}= {\bf f} \times {\bf f}^H$ to ${\tilde{\bf F}}\geq {\bf f}\times{\bf f}^H$
and applying the Schur complement theorem,
we get the following optimization problem:
\begin{eqnarray}\label{eqIII-B15}
       \min_{{\tilde{\bf F}},{\bf f}} && {\rm Tr}\left( {\bf Q}_0 {\tilde{\bf F}}\right) - {\bf f}^H {\bf q}_0 -{\bf q}^H_0 {\bf f} +q_0 \\ \nonumber
      s.t.  && {\rm Tr}\left( {\bf Q}_x {\tilde{\bf F}}\right) \leq P_R  \\ \nonumber
      && {\rm Tr}\left( {\bf Q}_k {\tilde{\bf F}}\right) \geq \lambda_k \sigma^2_k, \forall k \\ \nonumber
     && \left[\begin{array}{cc}
                      {\tilde{\bf F}} & {\bf f} \\
                      {\bf f}^H & 1
                      \end{array}\right]\geq 0
\end{eqnarray}
Note that \eqref{eqIII-B15} is convex, thus the obtained solution is optimal. If we generate enough samples of Gaussian variable ${\bf x}$ following ${\cal CN}( {\bar{{\bf f}}}, {\bar{{\tilde{\bf F}}}} - {\bar{{\bf f}}} \times {\bar{{\bf f}}}^H)$ with $\bar{{\tilde{\bf F}}}$ and $\bar{{\bf f}}$ being an optimal solution of \eqref{eqIII-B15},  and choose the best candidate $\bar {\bf x}$ from the samples as a solution of \eqref{eqIII-B44}, $\bar {\bf x}$ will optimally solve \eqref{eqIII-B44} on average, i.e.,
\begin{eqnarray}\label{eqIII-B16}
       \min_{{\bf f}} && {\cal E} \left( {\bf f}^H {\bf Q}_0 {\bf f} - {\bf f}^H {\bf q}_0 -{\bf q}^H_0 {\bf f} +q_0 \right)\\ \nonumber
      s.t.~ && {\cal E} \left( {\bf f}^H {\bf Q}_x {\bf f} \right) \leq P_R \\ \nonumber
     && {\cal E} \left( {\bf f}^H {\bf Q}_k {\bf f} \right) \geq \lambda_k \sigma^2_k, \forall k
\end{eqnarray}
Finally,
the proposed iterative algorithm for $K>2$ is outlined as:

\vspace{-0.2cm}
\hrulefill
\par
{\footnotesize
\textbf{Algorithm 2} (RS precoding with $K>2$)
\begin{itemize}
\item \textbf{Initialize} ${\bf F}$ 
\item \textbf{Repeat}
\begin{itemize}
\item Update the BS decoding matrix ${\bf W}$ using \eqref{eqnII-12} for a fixed ${\bf F}$;
\item Update the RS precoding matrix ${\bf F}$ with ${\bf W}$ fixed using the following steps: First, form an optimization problem as \eqref{eqIII-B10}, if the obtained $\bf F$ is rank-one, the optimal RS precoder is obtained by applying eigenvalue decomposition. Otherwise, apply the randomization procedures \eqref{eqIII-B13}-\eqref{eqIII-B15} to get a quasi-optimal solution;
\end{itemize}
\item \textbf{Until} termination criterion is satisfied.
\end{itemize}}
\vspace{-0.3cm}
\hrulefill

Note that although the obtained $\bf F$ from the second step in Algorithm 2 may not be optimal, our simulation results show that the obtained $\bf F$ by using randomization is always good enough to make the iteration convergent.

\subsubsection{Sum-rate maximization}
Motivated by the relationship between sum rate and weighted MMSE in MIMO-BC system recently found in \cite{Christensen2008},
we next try to extend the proposed RS precoding design for Total-MSE minimization to sum rate maximization.
The sum-rate maximization problem is re-stated as:
\begin{eqnarray} \label{eqIII-B16-1}
      \max_{{\bf F}} &&
      \log_2 \det \left({\bf I}_K + {\bf P}^H {\bf H}^H_2 {\bf F}^H  {\bf G}^H_1 \right. \\ \nonumber
      && \left. \left(\sigma^2_R {\bf G}_1 {\bf F} {\bf F}^H {\bf G}^H_1
      +   \sigma^2_B {\bf I}_N \right)^{-1}
      {\bf G}_1 {\bf F} {\bf H}_2{\bf P}\right) \\ \nonumber
      s.t. &&  \tau \leq P_R \\ \nonumber
      && \zeta_k \geq \lambda_k,~\forall k
\end{eqnarray}
where the constraints are the same with \eqref{eqIII-B1}. It is not hard to verify that \eqref{eqIII-B16-1} is non-convex. To solve \eqref{eqIII-B16-1}, we introduce the following lemma.

\textbf{Lemma 3}: If a $\bar{\bf F}$ satisfies the Karush-Kuhn-Tucker (KKT) conditions of \eqref{eqIII-B16-1}, it will also satisfy the KKT conditions of the following problem:
\begin{eqnarray} \label{eqIII-B16-2}
      \min_{{\bf F}} &&
      {\rm Tr}\left( {\bf A} {\bf E}^{-1} \right) \\ \nonumber
      s.t. &&  \tau \leq P_R \\ \nonumber
      && \zeta_k \geq \lambda_k,~\forall k
\end{eqnarray}
where ${\bf E}$ is defined in \eqref{eqnII-10}, if the weight matrix ${\bf A}$ is set to
\begin{equation}\label{eqIII-B16-3}
\begin{split}
{\bf A}=& \frac{1}{\log 2}\left( {\bf I}_{K} + {\bf P}^H {\bf H}^H_2 \bar{\bf F}^H  {\bf G}^H_1
\left(\sigma^2_R {\bf G}_1 \bar{\bf F} \bar{\bf F}^H {\bf G}^H_1 + \right. \right. \\
 & \left. \left. \sigma^2_B {\bf I}_N \right )^{-1}
      {\bf G}_1 \bar{\bf F} {\bf H}_2{\bf P} \right).
\end{split}
\end{equation}
\begin{proof}
The proof is similar to the MIMO BC precoding design problem in \cite{Christensen2008}, thus we omit for brevity.
\end{proof}
\textit{Lemma 3} implies that using the weight matrix $\bf A$ in \eqref{eqIII-B16-3}, \eqref{eqIII-B16-1} shares the same stationary point with \eqref{eqIII-B16-2}. Then alternating optimization can be used to get the final solution of \eqref{eqIII-B16-1} as in \cite{Christensen2008}, which is presented as follows:

\vspace{-0.2cm}
\hrulefill
\par
{\footnotesize
\textbf{Algorithm 3} (RS precoding for maximizing sum rate)
\begin{itemize}
\item \textbf{Initialize} ${\bf F}$ 
\item \textbf{Repeat}
\begin{itemize}
\item Update the BS decoder matrix ${\bf W}$ using \eqref{eqnII-12} for fixed ${\bf F}$ and ${\bf A}$;
\item Update the weight matrix ${\bf A}$ using \eqref{eqIII-B16-3} for fixed ${\bf F}$ and ${\bf W}$;
\item Update the RS precoder matrix ${\bf F}$ as in Algorithm 1 or 2;
\end{itemize}
\item \textbf{Until} termination criterion is satisfied.
\end{itemize}}
\vspace{-0.3cm}
\hrulefill

According to the convergence analysis provided in \cite{Christensen2008}, the convergence of Algorithm 3 can be ensured.

\subsection{Joint precoding}
Obviously, the previously presented two precoding designs can be combined to realize the joint BS-RS precoding design to obtain better performance. In this case, if the RS has enough capability to enable the joint design, it can collect all the required CSI and optimize $\bf F$ and $\bf B$ jointly. Then besides $\bf F$, the RS should also broadcast $\bf B$ to the BS and MSs. On the other hand,
the joint optimization can also be conducted at the BS
and the RS helps to collect CSI and transmits them to the BS.
Then, the BS needs to transmit $\bf B$ and $\bf F$ to the RS, and the RS further broadcasts them to the MSs.
Nevertheless, such joint precoding design requires more feedback overheads although it leads to better performance.

According to the algorithms proposed in Subsections A and B, the joint precoding design is outlined as:

\vspace{-0.2cm}
\hrulefill
\par
{\footnotesize
\textbf{Algorithm 4} (Joint precoding scheme)
\begin{itemize}
\item \textbf{Initialize} ${\bf B}$ 
\item \textbf{Repeat}
\begin{itemize}
\item Update the RS precoder $\bf F$ for a fixed BS precoder $\bf B$ by using Algorithm 1 or 2 for Total-MSE minimization and Algorithm 3 for sum rate maxmization;
\item Update the BS precoder $\bf B$ for a fixed relay precoder $\bf F$ by using the SOCP optimization as in Subsection A;
\end{itemize}
\item \textbf{Until} termination criterion is satisfied.
\end{itemize}}
\vspace{-0.3cm}
\hrulefill

\begin{table*}[t]
\centering
\caption{\label{Comparison} Signaling overhead and design complexity comparison}
\begin{tabular}{c|c|c|c|c|c}
\toprule
\multirow{2}{*}{ } & \multicolumn{2}{|c|}{TDD} & \multicolumn{2}{|c|}{FDD} & \multirow{2}{*}{Complexity} \\
\cline{2-5}
  & Overhead-I & Overhead-II & Overhead-I  & Overhead-II & \\
\hline
(1)\begin{tabular}{c}
                             BS Precoding \\
                             ( Design at BS ) \\
                           \end{tabular} &  \begin{tabular}{c}
                             RS $\stackrel{{\bf h}_{2k}\forall k}{\Longrightarrow}$ BS \\
                             RS $\stackrel{{\bf H}_1}{\Longrightarrow}$ MSs \\
                           \end{tabular} & \begin{tabular}{c}
                             BS $\stackrel{{\bf B},\alpha}{\Longrightarrow}$ RS \\
                             RS $\stackrel{{\bf B},\alpha}{\Longrightarrow}$ MSs \\
                           \end{tabular} & \begin{tabular}{c}
                          MSs $\stackrel{{\bf g}_{2k}\forall k}{\Longrightarrow}$ RS \\
                             RS $\stackrel{{\bf h}_{2k},{\bf g}_{2k}\forall k,{\bf H}_1}{\Longrightarrow}$ BS \\
                             RS $\stackrel{{\bf H}_1,{\bf h}_{2k}}{\Longrightarrow}$ MS $k$ \\
                           \end{tabular} & \begin{tabular}{c}
                             BS $\stackrel{{\bf B},\alpha}{\Longrightarrow}$ RS \\
                             RS $\stackrel{{\bf B},\alpha}{\Longrightarrow}$ MSs \\
                           \end{tabular} & $ O( n_{BS}  )$ \\
\hline
(2) \begin{tabular}{c}
                             BS Precoding \\
                             ( Design at RS ) \\
                           \end{tabular} & same as (1)  & RS $\stackrel{{\bf B},\alpha}{\Longrightarrow}$ BS, MSs & \begin{tabular}{c}
                           BS $\stackrel{{\bf G}_1}{\Longrightarrow}$ RS \\
                           MSs $\stackrel{{\bf g}_{2k}\forall k}{\Longrightarrow}$ RS \\
                             RS $\stackrel{{\bf h}_{2k}\forall k,{\bf H}_1}{\Longrightarrow}$ BS \\
                             RS $\stackrel{{\bf H}_1,{\bf h}_{2k}}{\Longrightarrow}$ MS $k$ \\
                           \end{tabular} & RS $\stackrel{{\bf B},\alpha}{\Longrightarrow}$ BS, MSs & $O(n_{BS})$ \\
\hline
(3) \begin{tabular}{c}
                             RS Precoding \\
                             ( Design at BS ) \\
                           \end{tabular} &  same as (1) & \begin{tabular}{c}
                             BS $\stackrel{{\bf F}}{\Longrightarrow}$ RS \\
                             RS $\stackrel{{\bf F}}{\Longrightarrow}$ MSs \\
                           \end{tabular} & same as (1) & \begin{tabular}{c}
                             BS $\stackrel{{\bf F}}{\Longrightarrow}$ RS \\
                             RS $\stackrel{{\bf F}}{\Longrightarrow}$ MSs \\
                           \end{tabular} & $ O( n_{RS}  )$ \\
\hline
(4) \begin{tabular}{c}
                             RS Precoding \\
                             ( Design at RS ) \\
                           \end{tabular} & same as (1)  & RS $\stackrel{{\bf F}}{\Longrightarrow}$ BS, MSs & same as (2) & RS $\stackrel{{\bf F}}{\Longrightarrow}$ BS, MSs & $O(n_{RS})$ \\

\hline
(5) \begin{tabular}{c}
                             Joint Precoding \\
                             ( Design at BS ) \\
                           \end{tabular} &  same as (1)
 & \begin{tabular}{c}
                             BS $\stackrel{{\bf B},{\bf F}}{\Longrightarrow}$ RS \\
                             RS $\stackrel{{\bf B},{\bf F}}{\Longrightarrow}$ MSs \\
                           \end{tabular} &  same as (1) & \begin{tabular}{c}
                             BS $\stackrel{{\bf B},{\bf F}}{\Longrightarrow}$ RS \\
                             RS $\stackrel{{\bf B},{\bf F}}{\Longrightarrow}$ MSs \\
                           \end{tabular} & $O({l_{J}}(n_{BS}+n_{RS}))$ \\
\hline
(6)  \begin{tabular}{c}
                             Joint Precoding \\
                             ( Design at RS ) \\
                           \end{tabular} &same as (1) & RS $\stackrel{{\bf B},{\bf F}}{\Longrightarrow}$ BS, MSs & same as (2) & RS $\stackrel{{\bf B},{\bf F}}{\Longrightarrow}$ BS, MSs & $O({l_{J}}(n_{BS}+n_{RS}))$ \\
\bottomrule
\end{tabular}
\end{table*}

\textbf{Lemma 4}: The proposed joint precoding design algorithm is convergent.
\begin{proof}
For convenience of presentation, we take Total-MSE minimization as example. The proof can be easily extended to the case of sum rate maximization.
Firstly, for a fixed $\bf F$, updating $\bf B$ must decrease the Total-MSE at the BS by increasing $\alpha$ in \eqref{eqnIII-1}, otherwise, the BS precoder $\bf B$ should not be changed. Thus, we have
\begin{equation}\label{eqIII-B17}\nonumber
    e\left({\bf B}(n+1),{\bf F}(n)\right)\leq e\left({\bf B}(n),{\bf F}(n)\right),
\end{equation}
where $n$ denotes the iteration index. Then, we apply the proposed RS precoding design to update $\bf F$ by initializing ${\bf F}_0=\alpha {\bf F}(n)$. Since the proposed iterative RS precoding design algorithm decreases Total-MSE after each iteration, we have\footnote{On the case of solving \eqref{eqIII-B44} through randomization at $K\geq 3$, if we cannot find a solution decreasing the objective value in \eqref{eqIII-B44}, we can just set ${\bf F}(n+1)= \alpha{\bf F}(n)$.}
\begin{equation}\label{eqIII-B18} \nonumber
    e\left({\bf B}(n+1),{\bf F}(n+1)\right)\leq e\left({\bf B}(n+1),{\bf F}(n)\right).
\end{equation}
Therefore, we conclude that the joint precoding design algorithm is convergent.
\end{proof}

\begin{figure}[t]
  \centering
  \subfigure[$N=2, M=2, K=2$]{
    \label{CheckOpt:subfig:a} 
    \includegraphics[scale=0.60]{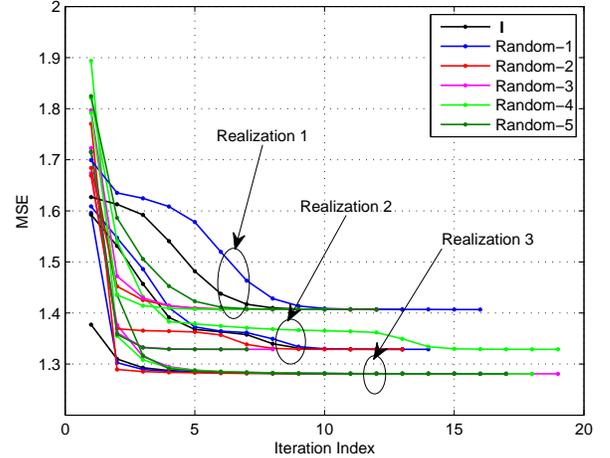}}
  \hspace{0in}
  \subfigure[$N=3, M=3, K=3$]{
    \label{CheckOpt:subfig:b} 
    \includegraphics[scale=0.60]{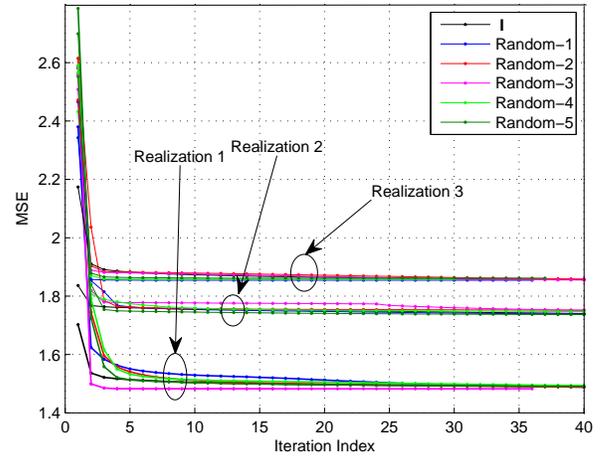}}
  \caption{Checking the optimality of the RS precoding design at $P=5$ dB and $L=5$.}
  \label{CheckOpt:subfig} 
\end{figure}

\section{Discussion on Signaling Overhead and design complexity}
As mentioned previously, each precoding design has its own merit. Choosing which precoding scheme is not only dependent on the processing capability of the BS and the RS, but also the design complexity and signaling overhead.
In this section, we provide a comprehensive comparison between these designs.
It is assumed that the channel characteristics of each link change slowly enough so that
they can be perfectly estimated by using pilot symbols or training sequences. Besides, the information of channel state and precoders can be exchanged accurately between the BS and the RS, the RS and the MSs through some lower rate auxiliary channels.
For completeness, two transmission modes, i.e., time-division duplex (TDD) mode and frequency-division duplex (FDD) mode, are considered, respectively.
The overall comparisons are presented in Table~\ref{Comparison}, where ``Overhead-I" denotes the overhead
used to feed back the CSI and ``Overhead-II" denotes the overhead used to feed back the precoding information.
Moreover, we suppose that the BS and MSs can estimate their local CSI ${\bf G}_1$ and ${\bf h}_{2k}$, $\forall k$, respectively.

Since the BS precoding design is a SOCP problem, according to \cite{SCOP}, the design complexity can be approximated as
\begin{equation}\label{eqIII-B18-1}
    n_{BS} = (NK+1)^2 (K+2)^{0.5} (2NK+K^2+2K+4)\log(1/\epsilon),
\end{equation}
where $\epsilon$ denotes the solution accuracy.
For the RS precoding design,
the design complexity mainly comes from solving the SDP problem and using the randomization technique.
Thus, according to \cite{Luo2010}, it can be approximated as
\begin{equation}\label{eqIII-B18-2}
    n_{RS} = l_{RS}\left(\max(M^2,K+2)^4 M \log(1/\epsilon)+ n_{rd}\right),
\end{equation}
where $n_{rd}$ denotes the complexity of randomization and $l_{RS}$ denotes the iteration number required in Algorithm 1, 2 or 3. Note that when $K \leq 2$, $n_{rd}$ is equal to $0$ (assuming that the complexity of getting rank-one solution from higher rank one can be omitted).
Combining \eqref{eqIII-B18-1} and \eqref{eqIII-B18-2}
leads to the joint precoding design complexity given in Table~\ref{Comparison} where $l_J$ denotes the iteration number needed in Algorithm 4.

From Table~\ref{Comparison}, we find that
the difference of signal overhead between the BS precoding and the RS precoding is not significant
if they are designed at the same station and it depends on the antenna configuration of the system. In general, the BS precoding design has less design complexity compared with the RS precoding design. For each precoding design, it is more practical to perform it at the RS in order to save the signaling overhead consumption.

\section{Simulation results}
In this section, some numerical examples are presented to evaluate the proposed precoding designs. The channels are set to be Rayleigh fading, i.e., the elements of each channel matrix or vector are complex Gaussian random variables with zero mean and unit variance. We assume that the noise powers at all the destinations are the same, i.e., $\sigma^2_B = \sigma^2_R =\sigma^2_k = 1$, $\forall k$. The transmission power at all the MSs and RS are the same as $P_R = P_k = P$, $\forall k$, and the transmission power at the BS is assumed to be $P_B = LP$ where $L$ is a constant.
For all the simulations, $1000$ channel realizations have been simulated. Moreover, $10000$ quadrature-phase-shift keying (QPSK) symbols are transmitted from each source node for each channel realization when simulating bit-error-rate (BER) performance.
For all comparisons, if not specified otherwise, the fixed RS procoder $\tilde{{\bf F}}$ in the BS precoding design is chosen as $\tilde{{\bf F}} = {\bf I}_M$ and the fixed BS precoder ${\bf B}$ in the RS precoding design is chosen as ${\bf B} = \sqrt{P_B/K} {\bf I}_{N \times K}$.

In Fig.~\ref{CheckOpt:subfig}, we check the optimality of the proposed RS precoding design, Algorithm 1 and Algorithm 2, for Total-MSE minimization in Fig.~\ref{CheckOpt:subfig:a} and Fig.~\ref{CheckOpt:subfig:b}, respectively, by trying different initialization points at three sets of given but arbitrary channel realizations. Specifically, for each channel realization, six different initialization points, including the identity matrix and five random matrices, are simulated. Moreover, for $K=3$, we choose three channel realizations where the randomization technique is needed to find a quasi-optimal rank-one solution of \eqref{eqIII-B44}.
Fig.~\ref{CheckOpt:subfig:a} shows that Algorithm 1 for $K=2$ can converge to a unique solution with any initialization points. Fig.~\ref{CheckOpt:subfig:b} shows that Algorithm 2 for $K>2$ is also able to converge to the solutions which are close to each other with different initialization points. Thus we conclude that the proposed iterative RS precoding for Total-MSE minimization can indeed approach the optimal solution.

\begin{figure}[tbhp]
  \centering
  \subfigure[Convergence behavior]{
    \label{Coverg:subfig:a} 
    \includegraphics[scale=0.60]{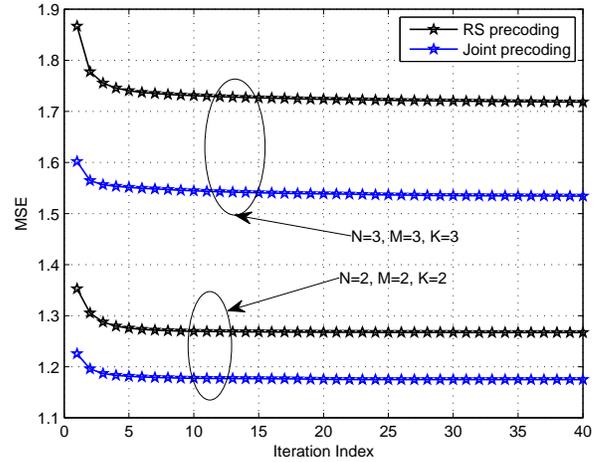}}
  \hspace{0in}
  \subfigure[Complexity of randomization]{
    \label{Coverg:subfig:b} 
    \includegraphics[scale=0.60]{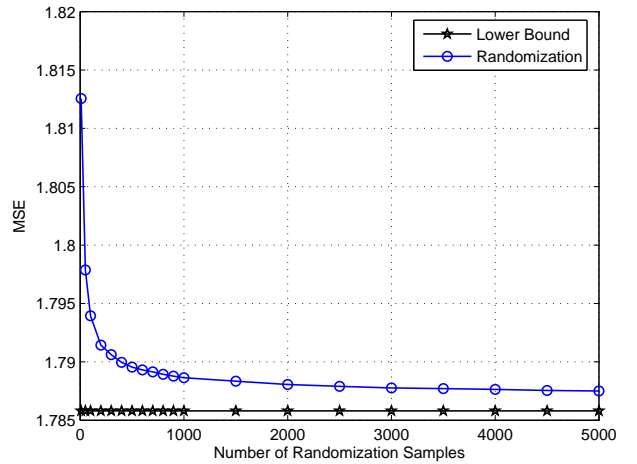}}
  \caption{Convergence behavior of the proposed iterative precoding design and complexity of randomization at $P=5$ dB and $L=5$.}
  \label{Coverg:subfig} 
\end{figure}

In Fig.~\ref{Coverg:subfig:a}, the convergence behavior of the proposed RS and joint precoding designs for Total-MSE minimization is shown as the function of iteration index at $P=5$ dB and $L=5$. We observe that the proposed RS precoding converges in $20$ iterations for $K=2$ and in $30$ iterations for $K=3$. Moreover, the proposed joint precoding algorithm converges within $10$ iterations for both two and three MSs\footnote{Here, for the inner RS precoding design, we set the maximum iteration number as $20$ for $K=2$ and $30$ for $K=3$.}. Fig.~\ref{Coverg:subfig:b} illustrates the required random samples in solving \eqref{eqIII-B44} by using randomization to approach the lower bound obtained from \eqref{eqIII-B15}. We observe that as the number of the samples increases, a better solution can be obtained. But when the number exceeds $2000$, the obtained solution does not change much, which further indicates that $2000$ samples are enough in general to generate a near optimal solution.

\begin{figure}[t]
  \centering
  \subfigure[BER comparison]{
    \label{CompareAlltwoL1:subfig:a} 
    \includegraphics[scale=0.60]{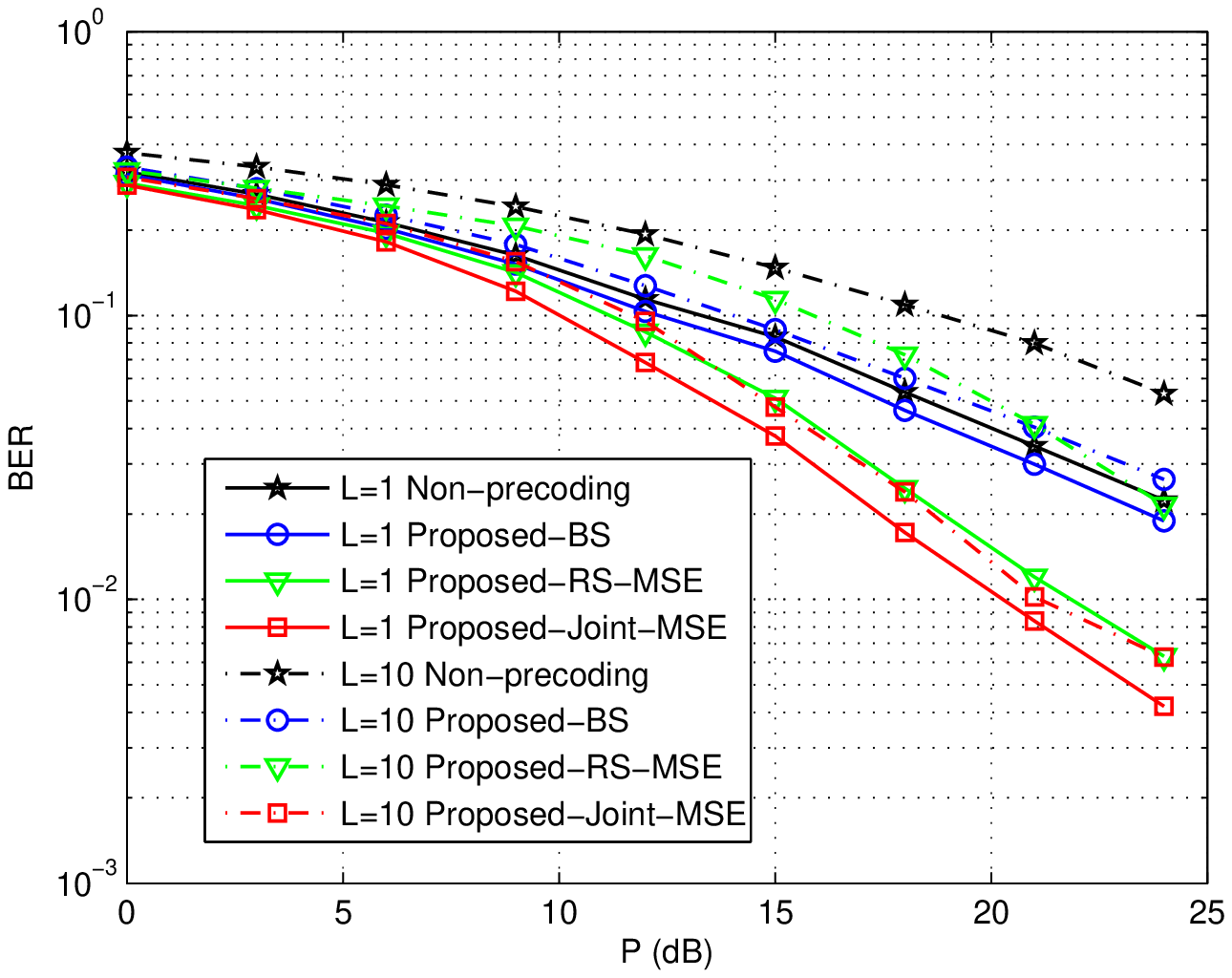}}
  \hspace{0in}
  \subfigure[Sum-rate comparison]{
    \label{CompareAlltwoL1:subfig:b} 
    \includegraphics[scale=0.60]{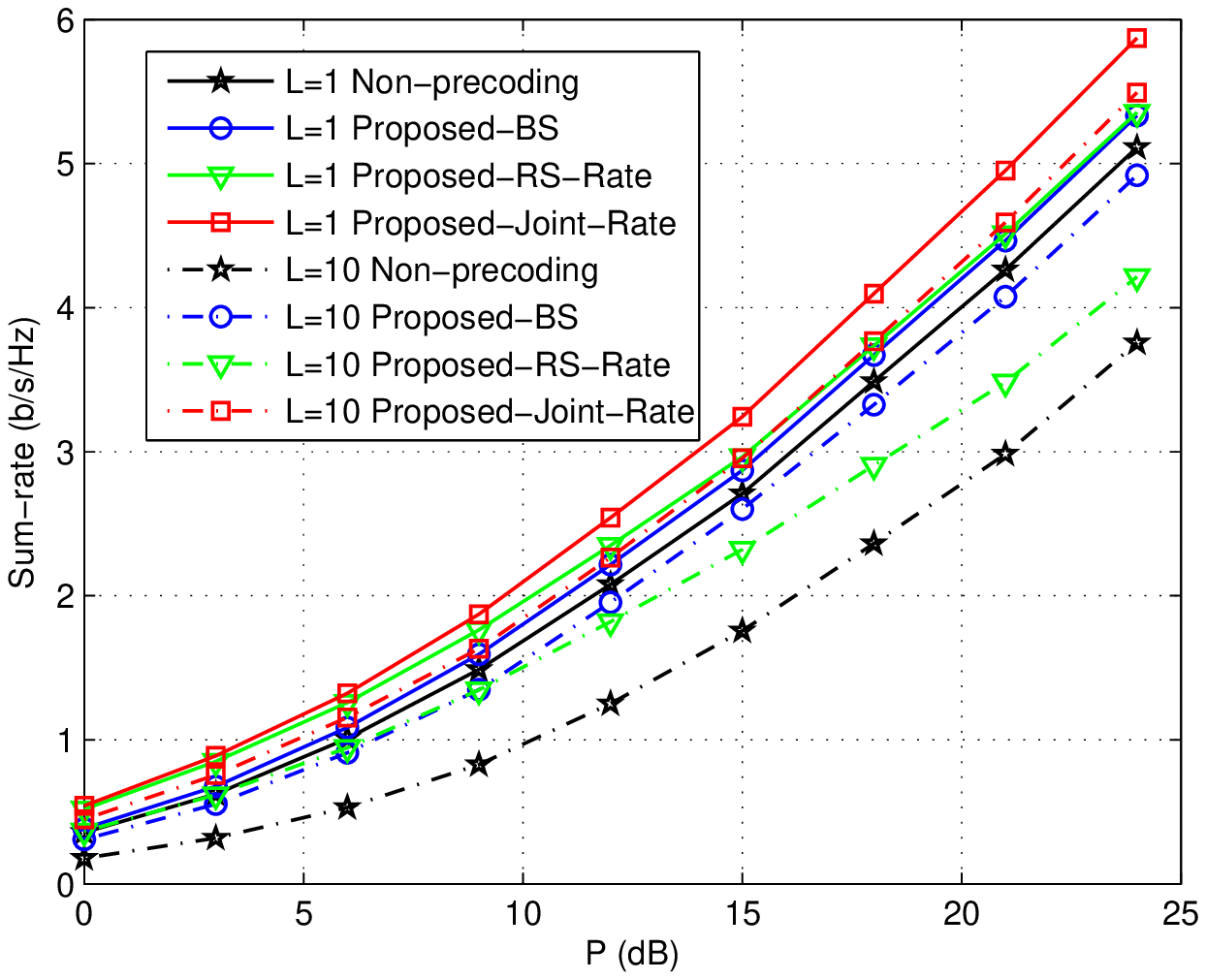}}
  \caption{Performance comparison for different precoding designs with $N=2, M=2, K=2$.}
  \label{CompareAlltwo:subfig} 
\end{figure}

In Fig.~\ref{CompareAlltwo:subfig}, we show the uplink BER and sum rate comparisons of all the proposed precoding designs as the function of $P$ for $N=2,M=2,K=2$ at $L=1$ and $L=10$ dB.
Here the notation ``-MSE" means that the precoding is designed based on the Total-MSE criterion, while ``-Rate" means that the precoding is designed based on the sum rate criterion.
For fair comparison and to make our optimization problems feasible,
we set the SINR requirements in \eqref{eqnII-13} as $\lambda_k=\epsilon_k,\forall k$ where $\epsilon_k$ is the SINR at the MS $k$ when no precoding is employed, i.e., both ${\bf B}$ and ${\bf F}$ are identity matrices.
We observe that when the BS has the same power as the RS and MS, i.e., $L=1$, the RS precoding design outperforms the the BS precoding design for both BER and sum rate comparison.
When the BS has more power than the RS and MS, i.e., $L=10$, the BS precoding can achieve better uplink performance than the RS precoding in certain SNR regime.
The reason is that with more power at the BS, the interference observed at each MS is introduced mainly by the downlink transmission. Then the precoding at the BS becomes important in coordinating the interference, which makes the BS precoding more effective than the RS precoding to improve the uplink performance.

\begin{figure}[t]
  \centering
  \subfigure[BER comparison]{
    \label{BSvsRS:subfig:a} 
    \includegraphics[scale=0.60]{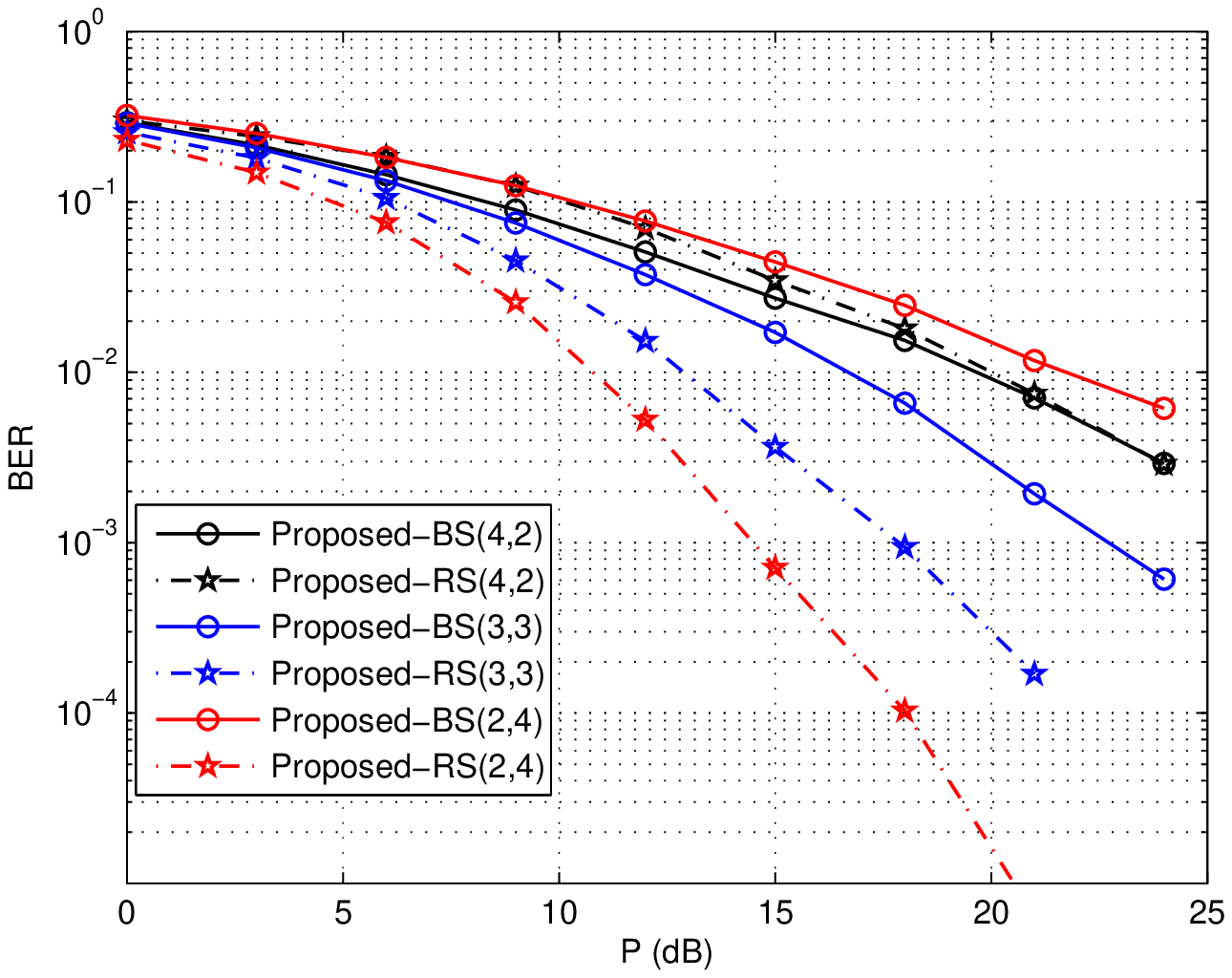}}
  \hspace{0in}
  \subfigure[Sum-rate comparison]{
    \label{BSvsRS:subfig:b} 
    \includegraphics[scale=0.60]{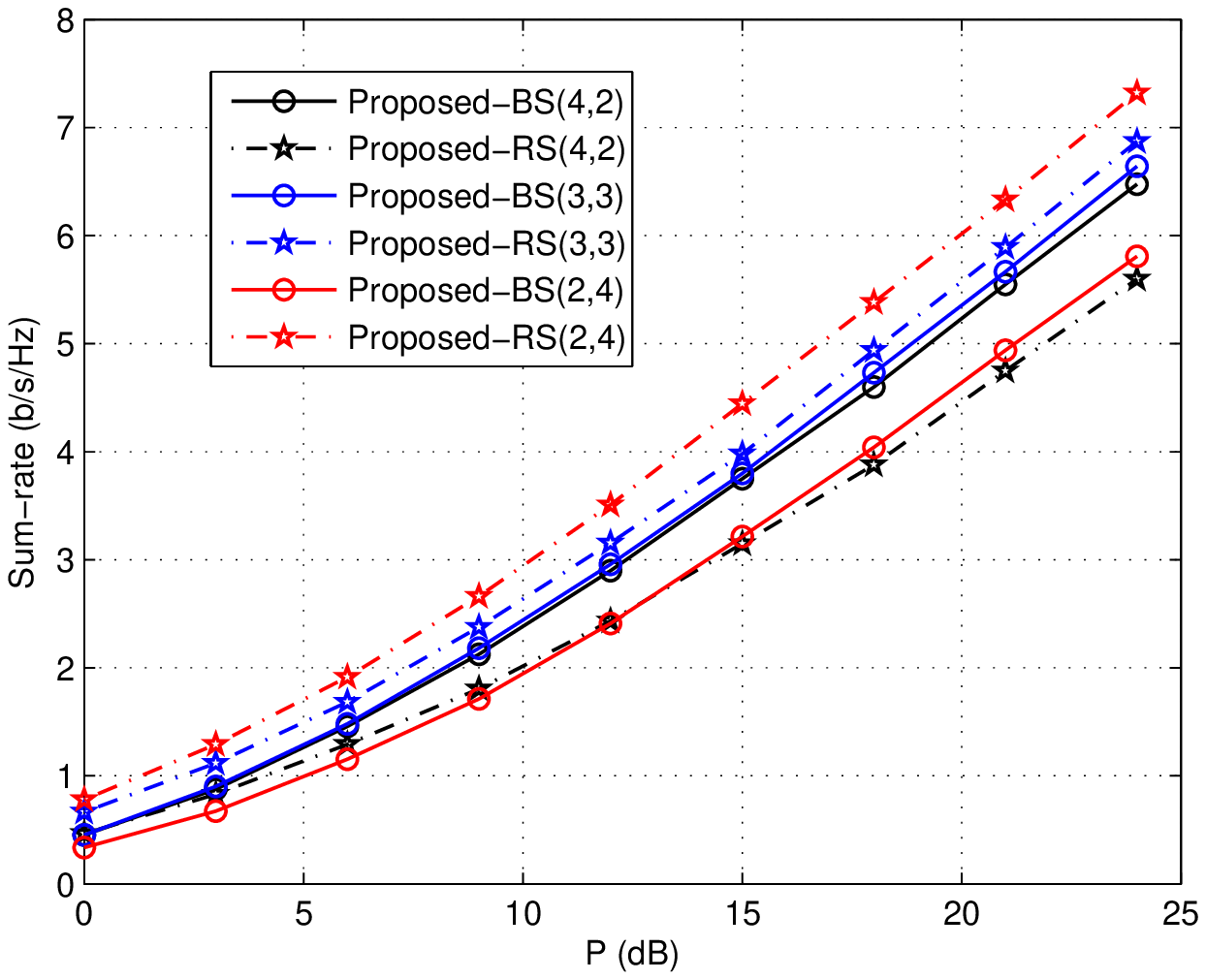}}
  \caption{Performance comparison for the BS and RS precoding designs with different antenna configuration $(N,M)$ at $L=5$ with $K=2$.}
  \label{BSvsRS:subfig} 
\end{figure}

\begin{figure}[t]
  \centering
  \subfigure[BER comparison]{
    \label{fig5:subfig:a} 
    \includegraphics[scale=0.60]{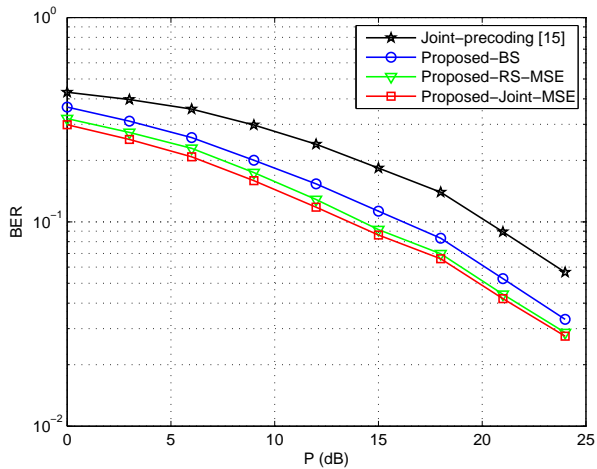}}
  \hspace{0in}
  \subfigure[Sum-rate comparison]{
    \label{fig5:subfig:b} 
    \includegraphics[scale=0.60]{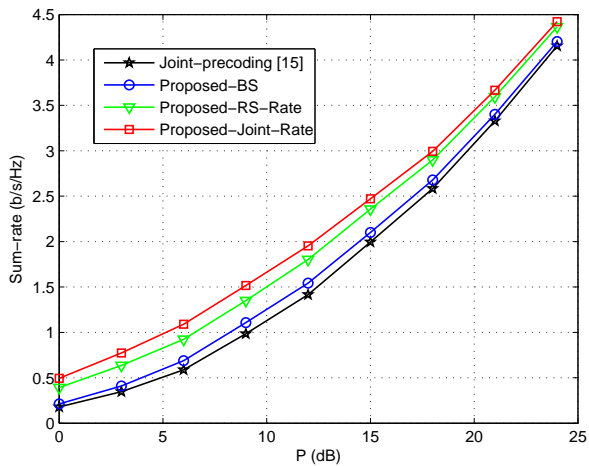}}
  \caption{Performance comparison with \cite{Ding2011} at $N=2, M=2, K=2$ and $L=10$.}
  \label{fig5:subfig} 
\end{figure}

Fig.~\ref{BSvsRS:subfig} illustrates the BER and sum rate comparison for different BS and RS antenna configuration $(N,M)$ at $K=2$ with total number of the BS and RS antennas being fixed at $N+M=6$.
For fair comparison, the target SINR at each MS is set as $\lambda_k=-5$dB, $\forall k$ and the uplink performance is averaged over the cases where the BS and RS precoding designs are feasible.
We see that
when the BS has more antennas than the RS,
i.e., at $(4,2)$, the BS precoding performs better than the RS precoding.
The reason is that increasing the number of the BS antennas is not only helpful for the BS precoding, but also helpful for the decoding of the uplink transmission.
However,
when the RS has more antennas than the BS,
the system performance can be significantly enhanced and the RS precoding greatly outperforms the BS precoding. This indicates that the antennas are more useful at the RS, while not at the BS. This is because
the BS precoding just makes an effort to let the downlink use less RS power to satisfy the SINR requirements at the MSs, and then more RS power can be allocated for
the uplink to improve the performance. However, the RS precoding is directly relevant to the uplink transmission. A well designed RS precoder can change the uplink channel matrix, not only the power.

In Fig.~\ref{fig5:subfig}, we compare the proposed precoding designs with the joint precoding design in \cite{Ding2011} for $K=2$ at $L=10$. For fairness, we set the SINR requirements in \eqref{eqnII-13} as $\lambda_k=\epsilon_k,\forall k$ where $\epsilon_k$ is the SINR obtained by using the precoders obtained in \cite{Ding2011}.
Specifically, the RS and BS precoders obtained from \cite{Ding2011} are chosen as the fixed RS precoder in ``Proposed-BS" and the fixed BS precoder in ``Proposed-RS", respectively.
Under this setup, we find that further optimizing the BS precoder or the RS precoder can obtain more performance gain over \cite{Ding2011}. Fig.~\ref{fig5:subfig} also shows that the RS precoding can get most of the performance gain of the joint precoding, which implies that the obtained ZF BS precoding in \cite{Ding2011} is indeed a good choice for improving the system performance.

\section{Conclusions}
In this paper, we studied linear precoding designs for multiuser two-way relay systems in a cellular network
for maximizing the uplink performance while maintaining the downlink QoS requirements.
Three precoding schemes were considered, namely, the BS precoding, the RS precoding and the joint BS-RS precoding.
By recasting the precoding designs into suitable forms, we obtained the optimal solution for the BS precoding and the local optimal solutions for both the RS precoding and the joint BS-RS precoding.
The performance of these precoding designs were compared and some practical implementation issues were discussed.
Simulation results showed that the RS precoding design is more efficient than the BS precoding design in most cases.
The results also demonstrated the superiority  of the proposed precoding designs over existing ones.

\appendices
\section{Proof of lemma 1}
\label{prof_lemma1}
To prove \textit{Lemma 1}, we only need to verify that functions $f_1 (\beta) = {\rm Tr}\left( {\bf E}(\beta)^{-1}\right)$ and $f_2 (\beta) = \log_2 \det \left( {\bf E}(\beta) \right)$ with
\begin{equation}\nonumber
\begin{split}
     {\bf E}(\beta) = & {\bf I}_{K} +  {\bf P}^H {\bf H}^H_2 \tilde{{\bf F}}^H{\bf G}^H_1 \\
      &  \left(\sigma^2_R {\bf G}_1\tilde{{\bf F}}\tilde{{\bf F}}^H {\bf G}^H_1 +\beta \sigma^2_B {\bf I}_{N} \right)^{-1}
      {\bf G}_1 \tilde{{\bf F}} {\bf H}_2{\bf P},
\end{split}
\end{equation}
where $\beta=1/\alpha^2$, are monotonically increasing and decreasing with respect to $\beta$, respectively.
To this end, we have
\begin{equation}\nonumber
\begin{split}
      & \frac{df_1(\beta)}{d\beta}=
      {\rm Tr}\bigg(-{\bf E}(\beta)^{-1} \\
      & \frac{d \big({\bf P}^H {\bf H}^H_2 \tilde{{\bf F}}^H {\bf G}^H_1  ( {\sigma^2_R {\bf G}_1 \tilde{{\bf F}}\tilde{{\bf F}}^H{\bf G}^H_1 +\beta \sigma^2_B {\bf I}_{N}}  )^{-1} {\bf G}_1 \tilde{{\bf F}} {\bf H}_2{\bf P} \big) } {d \beta}  \\
      & {\bf E}(\beta)^{-1} \bigg )\\
      &={\rm Tr}\left(\sigma^2_B  {\bf E}(\beta)^{-1}{\bf P}^H {\bf H}^H_2 \tilde{{\bf F}}^H {\bf G}^H_1 {\bf R}^{-2}
      {\bf G}_1 \tilde{{\bf F}} {\bf H}_2 {\bf P}{\bf E}(\beta)^{-1}\right )\\
      & > 0,
\end{split}
\end{equation}
\begin{equation}\nonumber
\begin{split}
      & \frac{df_2(\beta)}{d\beta}=
      \frac{1}{\log 2} {\rm Tr}\bigg( {\bf E}(\beta)^{-1} \\
      & \frac{d \big({\bf P}^H {\bf H}^H_2 \tilde{{\bf F}}^H {\bf G}^H_1  ( {\sigma^2_R {\bf G}_1\tilde{{\bf F}}\tilde{{\bf F}}^H {\bf G}^H_1 +\beta \sigma^2_B {\bf I}_{N}}  )^{-1} {\bf G}_1 \tilde{{\bf F}} {\bf H}_2{\bf P} \big) } {d \beta}
       \bigg)\\
      &=-\frac{1}{\log 2} {\rm Tr}\left (\sigma^2_B  {\bf E}(\beta)^{-1}{\bf P}^H {\bf H}^H_2 \tilde{{\bf F}}^H {\bf G}^H_1 {\bf R}^{-2}
      {\bf G}_1 \tilde{{\bf F}} {\bf H}_2 {\bf P}\right )\\
      &< 0,
\end{split}
\end{equation}
where ${{\bf R}} = {\sigma^2_R {\bf G}_1 \tilde{{\bf F}}\tilde{{\bf F}}^H {\bf G}^H_1 +\beta \sigma^2_B {\bf I}_{N}} $. For both inequalities, we have used the fact that both ${\bf E}$ and ${\bf R}$ are positive definite.
Thus, the proof is completed.

\section{Transformations from \eqref{eqIII-B4} to \eqref{eqIII-B44}}
\label{transformation}
We first rewrite the objective function $f({\bf F},{\bf W})$ in  \eqref{eqIII-B4} as
\begin{equation} \label{eqIII-B4-1}
       f({\bf F},{\bf W})  = {\bf f}^H {\bf Q}_0 {\bf f} - {\bf f}^H {\bf q}_0 -{\bf q}^H_0 {\bf f} +q_0,
\end{equation}
where ${\bf q}_0=vec({\bf G}^H_1 {\bf W}^H_1 {\bf P}^H {\bf H}^H_2)$, $q_0={\rm Tr}\left(\sigma^2_B {\bf W} {\bf W}^H + {\bf I}_{K}\right)$ and
\begin{equation}\label{eqIII-B5}
\begin{split}
       {\bf f} &=vec({\bf F}), \\
      {\bf Q}_0  &= \left({\bf H}_2 {\bf P}  {\bf P}^H {\bf H}^H_2+\sigma^2_R {\bf I}_{M}\right)^T\otimes \left({\bf G}^H_1 {\bf W}^H {\bf W} {\bf G}_1\right).
\end{split}
\end{equation}
Here the second and third terms in \eqref{eqIII-B4-1} are obtained from the corresponding terms of the objective function in \eqref{eqIII-B4} by using the rule ${\rm Tr}({\bf A}^T {\bf B}) = (vec({\bf A}))^T vec({\bf B})$ \cite{xiandazhang2004}. The first term of \eqref{eqIII-B4-1} is the reformulation of the first term of the objective function in \eqref{eqIII-B4} by using the rule \cite{xiandazhang2004}
\begin{equation}\label{eqIII-B5-1}
     {\rm Tr}\left( {\bf A} {\bf B} {\bf C} {\bf D}\right)=\left(vec({\bf D}^T)\right)^T \left({\bf C}^T \otimes {\bf A}\right) vec({\bf B}).
\end{equation}
Again according to \eqref{eqIII-B5-1}, the relay power constraint $\tau \leq P_R$ in \eqref{eqIII-B4} can be re-expressed as
\begin{equation}\label{eqIII-B5-2} \nonumber
{\bf f}^H {\bf Q}_x {\bf f} \leq P_R,
\end{equation}
where
\begin{equation}\label{eqIII-B5-3}
     {\bf Q}_x = \left({\bf H}_1 {\bf B}{\bf B}^H {\bf H}^H_1 + {\bf H}_2 {\bf P} {\bf P}^H {\bf H}^H_2 + \sigma^2_R {\bf I}_{M}\right)^T\otimes {\bf I}_{M}.
\end{equation}
The SINR constraint $\zeta_k \geq \lambda_k$ in \eqref{eqIII-B4} is equivalent to, by simple manipulations
\begin{equation}\label{eqIII-B5-5}
\begin{split}
      & {\rm Tr} \bigg( {\bf g}^{*}_{2k}{\bf g}^T_{2k} {\bf F}  \bigg({\bf H}_1 {\bf b}_k {\bf b}^H_k {\bf H}^H_1-  \\
      & \lambda_k \big( \sum_{i\neq k}( {\bf H}_1 {\bf b}_i {\bf b}^H_i {\bf H}^H_1 + P_i {\bf h}_{2i} {\bf h}^H_{2i}) + \sigma^2_R {\bf I}_{M} \big)\bigg){\bf F}^H \bigg) \geq \lambda_k \sigma^2_k.
\end{split}
\end{equation}
By using \eqref{eqIII-B5-1}, inequality \eqref{eqIII-B5-5} can be rewritten as
\begin{equation}\label{eqIII-B5-6} \nonumber
{\bf f}^H {\bf Q}_k {\bf f} \geq \lambda_k \sigma^2_k,
\end{equation}
where
\begin{equation}\label{eqIII-B7}
\begin{split}
      & {\bf Q}_k = \bigg({\bf H}_1 {\bf b}_k {\bf b}^H_k {\bf H}^H_1- \\
      & \lambda_k \big( \sum_{i\neq k}({\bf H}_1 {\bf b}_i {\bf b}^H_i {\bf H}^H_1 + P_i {\bf h}_{2i} {\bf h}^H_{2i}) + \sigma^2_R {\bf I}_{M} \big)\bigg)^T
      \otimes \left({\bf g}^*_{2k}{\bf g}^T_{2k}\right).
\end{split}
\end{equation}
Finally, \eqref{eqIII-B4} can be readily written into a form as \eqref{eqIII-B44}.

\section{Proof of Theorem 1}
\label{prof_Theorem1}
Note that if $K=1$, the optimal rank-one solution can be obtained as claimed in
\textit{Lemma 3.1} given in \cite{Huang2010}, here we omit it for brevity. On the case where \eqref{eqIII-B10} has an optimal rank-one solution, it is indeed the optimal solution of \eqref{eqIII-B9}.
Next we focus on the case $K=2$ and the rank of the optimal solution of \eqref{eqIII-B10} is higher than one. Since the optimization problem \eqref{eqIII-B10} is convex,
the sufficient and necessary optimality conditions (or termed as complementary slackness condition) are
\begin{equation}\label{App-B1}
\begin{split}
       &y_k {\bf Tr}\left({ {\tilde{\bf Q}_k}{\bf X}_F}\right)=0,~ y_k \geq 0,k=1,2\\
       &y_3 {\bf Tr}\left({ {\tilde{\bf Q}_x}{\bf X}_F}\right)=0,~
       y_4 \left({\bf Tr}\left({{{\bf Q}}{\bf X}_F}\right)-1\right)=0,\\
       & y_3 \geq 0, y_4 \in \mathbb{R}
\end{split}
\end{equation}
where $y_i$, for $i=1,2,3,4$, are dual variables and
\begin{equation}\label{App-B2}
       {\rm Tr}\left({\bf Z}{\bf X}_F\right)=0
\end{equation}
with ${\bf Z}={\tilde{\bf Q}_0}+ y_1{\tilde{\bf Q}_1} +y_2{\tilde{\bf Q}_2}+y_3{\tilde{\bf Q}_x} +y_4{{\bf Q}} \succeq 0$.
To proceed, we assume that $\alpha_i= {\rm Tr}({\tilde{\bf Q}_i} {\bf X}_F ),i=1,2,x$.

We first consider the case where at least one inequality constraint in \eqref{eqIII-B10} is inactive, i.e., at least one $\alpha_i<0$.
Suppose that the rank of the obtained ${\bf X}_F$ in \eqref{eqIII-B10} is $R$ and it can be decomposed as ${\bf X}_F={\bf V}{\bf V}^H$ with ${\bf V}\in {\mathbb C}^{(M^2+1) \times R}$.
By applying the trick used in \cite{Huang2010}, we introduce a Hermitian matrix ${\bf M}$ to satisfy
\begin{equation}\label{App-B3}
       {\bf Tr}\left({\bf V}^H{\tilde{\bf Q}_k}{\bf V} {\bf M}\right)=0,~{\bf Tr}\left({\bf V}^H{\tilde{\bf Q}_x}{\bf V} {\bf M}\right)=0,~k=1,2,
\end{equation}
where ${\bf M}\in {\mathbb C}^{R \times R}$ has $R^2$ real elements. If $R^2\geq 3$, there always exists a nonzero solution ${\bf M}$ satisfying \eqref{App-B3}. Let $\delta_i$, for $i=1,2,\ldots,R$, be the eigenvalues of $\bf M$ and define $|\delta_0|=\max \{|\delta_i|,\forall i\}$. Then, we get ${\bf X}^{'}_F={\bf V}\left({\bf I}_R -(1/\delta_0){\bf M}\right){\bf V}^H$ and further set ${\bf X}^{''}_F={\bf X}^{'}_F/a$ with $a={\bf X}^{'}_F(1,1)$.
Here we note that $a>0$ due to the fact that ${\bf X}^{'}_F$ is positive semidefinite and ${\bf Q}_k$ is positive definite.
It is not hard to see that the rank of ${\bf X}^{''}_F$ is reduced by at least one. We next verify that ${\bf X}^{''}_F$ is still an optimal solution of \eqref{eqIII-B10}. First, we check the primal feasibility of ${\bf X}^{''}_F$. With ${\bf X}^{''}_F(1,1)=1$, the condition ${\bf Tr}\left({\bf Q}{\bf X}^{''}_F\right)=1$ is satisfied. Moreover, since ${\bf Tr}\left({\tilde{\bf Q}_i}{\bf X}^{'}_F\right)={\bf Tr}\left({\tilde{\bf Q}_i}{\bf X}_F\right),i=1,2,x$ and $a>0$, ${\bf Tr}\left({\tilde{\bf Q}_x}{\bf X}^{''}_F\right)\leq 0$ and ${\bf Tr}\left({\tilde{\bf Q}_i}{\bf X}^{''}_F\right)\leq 0,i=1,2$ are also satisfied. Second, we need to check the complementary conditions in \eqref{App-B1} and \eqref{App-B2}. It is found that if ${\bf Tr}\left({\tilde{\bf Q}_i}{\bf X}_F\right)=0,i=1,2,x$, we must have ${\bf Tr}\left({\tilde{\bf Q}_i}{\bf X}^{'}_F\right)=0,i=1,2,x$. Then ${\bf Tr}\left({\tilde{\bf Q}_i}{\bf X}^{''}_F\right)=0$, $i=1,2,x$, succeed, which means that \eqref{App-B1} is satisfied. On the other hand, if ${\bf Tr}\left({\tilde{\bf Q}_i}{\bf X}_F\right)\neq0$, $i=1,2,x$, it means that $y_i=0$. Then dividing ${\bf X}^{'}_F$ by $a$ does not affect the satisfaction of \eqref{App-B1}. For \eqref{App-B2}, since ${\rm Tr}\left({\bf Z}{\bf X}^{'}_F\right)={\rm Tr}\left({\bf Z}{\bf X}_F\right)=0$, ${\rm Tr}\left({\bf Z}{\bf X}^{''}_F\right)$ must be equal to zero. Thus ${\bf X}^{''}_F$ also satisfies the condition in \eqref{App-B2}. Therefore, ${\bf X}^{''}_F$ is also an optimal solution of \eqref{eqIII-B10}. Repeat the above procedure until $R^2\leq 3$, then an optimal rank-one solution is obtained. For completeness, we present the detailed procedures as follows:
\vspace{0.1cm}
\par
{\footnotesize
\begin{itemize}
\item Solve optimization problem \eqref{eqIII-B10} and get the optimal solution ${\bf X}_F$ with rank $R$;
\item \textbf{Repeat}
\begin{itemize}
\item Decompose ${\bf X}_F$ as ${\bf X}_F={\bf V}{\bf V}^H$;
\item Find a nonzero $R\times R$ Hermitian solution $\bf M$ of the following linear equations
       ${\bf Tr}\left({\bf V}^H{\tilde{\bf Q}_i}{\bf V} {\bf M}\right)=0,~i=1,2,x$;
\item Evaluate the eigenvalues $\delta_1,\delta_2,\cdots,\delta_R$ of $\bf M$ and set $|\delta_0|=\max \{|\delta_i|,\forall i\}$;
\item Compute ${\bf X}^{'}_F={\bf V}\left({\bf I}_R -(1/\delta_0){\bf M}\right){\bf V}^H$ and further get ${\bf X}^{''}_F={\bf X}^{'}_F/a$
with $a={\bf X}^{'}_F(1,1)$. 
\item Set ${\bf X}_F={\bf X}^{''}_F$.
\end{itemize}
\item \textbf{Until} the rank $R=\text{Rank}({\bf X}_F)$ is equal to $1$.
\end{itemize}}
\vspace{0.1cm}

Then we consider the case where all the inequality constraints are active, i.e., $\alpha_i = 0$, for $i =1,2,x$. Note that since $M^2+1>4$, the size of matrix ${\bf X}_F$ in \eqref{eqIII-B10} is always larger than four.
Suppose $R \geq 3$. Based on \emph{Theorem 2.1}
given in \cite{Wenbao2011}, we obtain that there is a rank-one decomposition for ${\bf X}_F$ (synthetically denoted as $\mathcal{D}_3 ({\bf X}_F, {\tilde{\bf Q}_1},{\tilde{\bf Q}_2},{\tilde{\bf Q}_x})$), i.e., ${\bf X}_F = \sum^R_{r=1} {\bf x}_r {\bf x}^H_r$, such that
\begin{equation}\label{App-C1}\nonumber
       {\bf x}^H_r {\tilde{\bf Q}_k} {\bf x}_r = \frac{{\rm Tr}({\tilde{\bf Q}_k} {\bf X}_F)}{R} = 0, ~k=1,2,~r=1,2,\cdots, R,
\end{equation}
and
\begin{equation}\label{App-C2}\nonumber
       {\bf x}^H_r {\tilde{\bf Q}_x} {\bf x}_r = \frac{{\rm Tr}({\tilde{\bf Q}_x} {\bf X}_F)}{R} = 0, ~r=1,2,\cdots, R-2.
\end{equation}
By generating ${\bf X}^{'}_F = {\bf x}_1 {\bf x}^H_1 $ and ${\bf X}^{''}_F = {\bf X}^{'}_F/{\bf X}^{'}_F(1,1)$ (we again note that ${\bf X}^{'}_F(1,1)>0$), it is easy to check ${\bf X}^{''}_F $ is feasible for \eqref{eqIII-B10} and satisfies the optimality conditions \eqref{App-B1} and \eqref{App-B2} together with the optimal dual solution $\{y_1, y_2,y_3,y_4\}$. Therefore, ${\bf X}^{''}_F$ can be regarded as an optimal rank-one solution of \eqref{eqIII-B10}.

\section{Procedure to get a suboptimal rank-one solution}
\label{procedure}
If \eqref{eqIII-B10} has an optimal solution of rank two with all the constraints being active,
we next give a method to obtain a good feasible solution.
Let the optimal solution in \eqref{eqIII-B10} be in the form ${\bf X}_F = \left[
                                                                                                                                                          \begin{array}{cc}
                                                                                                                                                            1 & {\bf z}^H \\
                                                                                                                                                             {\bf z}& {\bf X} \\
                                                                                                                                                          \end{array}
                                                                                                                                                        \right]
$. We have
\begin{equation}\label{App-C3}\nonumber
\begin{split}
      \alpha_k &= {\rm Tr}({\tilde{\bf Q}_k} {\bf X}_F) = \lambda_k \sigma^2_k - {\rm Tr}({\bf Q}_k {\bf X}), k=1,2,\\
      \alpha_x &= {\rm Tr}({\tilde{\bf Q}_x} {\bf X}_F) = -P_R + {\rm Tr}({\bf Q}_x {\bf X}).
\end{split}
\end{equation}
That is,
\begin{equation}\label{App-C4}\nonumber
\begin{split}
      \beta_k &= {\rm Tr}({\bf Q}_k {\bf X}) = \lambda_k \sigma^2_k - \alpha_k, k=1,2,\\
      \beta_x &= {\rm Tr}({\bf Q}_x {\bf X}) = \alpha_x + P_R.
\end{split}
\end{equation}
Then we have ${\rm Tr}( ({\bf Q}_k - \frac{\beta_k}{\beta_x} {\bf Q}_x) {\bf X} ) =0$. Again according to \emph{Theorem 2.1} given in \cite{Yongwei2007}, we obtain that there is a rank-one matrix decomposition (synthetically denoted as $\mathcal{D}_2 ({\bf X}_F, {\bf Q}_1 - \frac{\beta_1}{\beta_x} {\bf Q}_x,{\bf Q}_2 - \frac{\beta_2}{\beta_x} {\bf Q}_x)$) ${\bf X}=\sum^{\bar{R}}_{r=1} {\bf f}_r {\bf f}^H_r$ ($\bar{R}={\rm Rank}({\bf X}) \leq R$) such that
\begin{equation}\label{App-C5}\nonumber
      {\bf f}^H_r ({\bf Q}_k - \frac{\beta_k}{\beta_x} {\bf Q}_x) {\bf f}_r =  0, ~k=1,2,~r=1,2,\cdots, \bar{R}.
\end{equation}
We take ${\bf f}_1$ and set $\gamma = \frac{\beta_x}{{\bf f}^H_1{\bf Q}_x {\bf f}_1}$. It can be verified that $(\sqrt{\gamma}{\bf f}_1)^H {\bf Q}_k (\sqrt{\gamma}{\bf f}_1)=\beta_k$, for $k=1,2,x$, and that ${\bf X}^{'}={\bf x}_1 {\bf x}^H_1$ with ${\bf x}_1 = [1, (\sqrt{\gamma}{\bf f}_1)^H]^H$ is feasible for \eqref{eqIII-B10} and can be regarded as a suboptimal rank-one solution of \eqref{eqIII-B10}.

\bibliography{Journal_Multiuser}

\begin{IEEEbiography}[
]{Rui Wang}
received the B.S. degree from Anhui Normal University, Wuhu, China, in
2006, and the M.S. degree from Shanghai University, Shanghai, China, in 2009
both in electronic engineering.
Currently he is pursuing his Ph.D. degree at the Institute
of Wireless Communication Technology (IWCT) in
Shanghai Jiao Tong University. From August 2012 to February of 2013,
he is a visiting Ph.D student at the Department of Electrical Engineering of University of California, Riverside.

His research interests include digital image processing, cognitive radio and
signal processing for wireless cooperative communication.
\end{IEEEbiography}

\begin{IEEEbiography}[
]
{Meixia Tao}
(S'00-M'04-SM'10) received the B.S. degree in electronic engineering from Fudan University, Shanghai, China, in 1999, and the Ph.D. degree in electrical and electronic engineering from Hong Kong University of Science and Technology in 2003. She is currently an Associate Professor with the Department of Electronic Engineering, Shanghai Jiao Tong University, China. From August 2003 to August 2004, she was a Member of Professional Staff at Hong Kong Applied Science and Technology Research Institute Co. Ltd. From August 2004 to December 2007, she was with the Department of Electrical and Computer Engineering, National University of Singapore, as an Assistant Professor. Her current research interests include cooperative transmission, physical layer network coding, resource allocation of OFDM networks, and MIMO techniques.

Dr. Tao is an Editor for the \textsc{IEEE Transactions on Communications} and the \textsc{IEEE Wireless Communications Letters}. She was on the Editorial Board of the \textsc{IEEE Transactions on Wireless Communications} from 2007 to 2011 and the \textsc{IEEE Communications Letters} from 2009 to 2012. She also served as Guest Editor for \textsc{IEEE Communications Magazine} with feature topic on LTE-Advanced and 4G Wireless Communications in 2012, and Guest Editor for \textsc{EURISAP J WCN} with special issue on Physical Layer Network Coding for Wireless Cooperative Networks in 2010. She was in the Technical Program Committee for various conferences, including IEEE INFOCOM, IEEE GLOBECOM, IEEE ICC, IEEE WCNC, and IEEE VTC.

Dr. Tao is the recipient of the IEEE ComSoC Asia-Pacific Outstanding Young Researcher Award in 2009.
\end{IEEEbiography}

\begin{IEEEbiography}[
]{Yongwei Huang}
(M'09) received the Bachelor of Science degree in information and computation science in 1998 and the Master of Science degree in operations research in 2000, both from Chongqing University. In 2005, he received the Ph.D. degree in operations research from Chinese University of Hong Kong.

He is a Research Assistant Professor in Department of Mathematics, Hong Kong Baptist University, Hong Kong, which he joined in 2011. Prior to the current position, he had held several research appointments in Department of Biomedical, Electronic, and Telecommunication Engineering, University of Naples "Federico II," Italy; Department of Electronic and Computer Engineering, Hong Kong University of Science and Technology; and Department of Systems Engineering and Engineering Management, Chinese University of Hong Kong.

His research interests are related to optimization theory and algorithms, including conic optimization, robust optimization, combinatorial optimization, and stochastic optimization, and their applications in signal processing for radar and wireless communications. He is a recipient of Best Poster Award in the 2007 Workshop on Optimization and Signal Processing.
\end{IEEEbiography}

\end{document}